\documentstyle[12pt,epsfig]{article}
\begin{document}

\begin{center}
{\bf HEAVY QUARK PRODUCTION IN HADRON COLLISIONS}

\vspace{0.5cm}

M.G.Ryskin, Yu.M.Shabelski and A.G.Shuvaev \\
Petersburg Nuclear Physics Institute, \\
Gatchina, St.Petersburg 188350 Russia \\

\end{center}

\vspace{0.5cm}
{\it Lecture, given at XXXIV PNPI Winter School, \\
St.Petersburg, 2000}

\vspace{0.5cm}

\begin{abstract}
We compare the numerical predictions for heavy quark production in high
energy hadron collisions of the conventional QCD parton model and the
$k_T$-factorization approach (semihard theory). The total production
cross sections, one-particle rapidity and $p_T$ distributions as well as
several two-particle correlations are considered. Some of them can
help to estimate QCD scale. The difference in the predictions of the
two approaches is not very large, while the shapes of the
distributions are slightly different.

\end{abstract}

\vspace{2cm}

E-mail RYSKIN@THD.PNPI.SPB.RU

E-mail SHABELSK@THD.PNPI.SPB.RU

E-mail SHUVAEV@THD.PNPI.SPB.RU

\newpage

\section{Introduction}

The investigation of heavy quark production in high energy hadron
collisions is an important method for studying the quark-gluon
structure of hadrons. Realistic estimates of the cross section of
heavy quark production as well as their correlations are necessary in
order to plan experiments on existing and future accelerators or
in cosmic ray physics.

The description of hard interactions in hadron collisions within the
framework of QCD is possible only with the help of some phenomenology,
which reduces the hadron-hadron interaction to the parton-parton one via
the formalism of the hadron structure functions. The cross sections of
hard processes in hadron-hadron interactions can be written as the
convolutions squared matrix elements of the sub-process calculated
within the framework of QCD, with the parton distributions in the
colliding hadrons.

The most popular and technically simplest approach is the so-called QCD
collinear approximation, or parton model (PM). In this model all
particles involved are assumed to be on mass shell, carrying only
longitudinal momenta, and the cross section is averaged over two
transverse polarizations of the incident gluons. The virtualities $q^2$
of the initial partons are taken into account only through their
structure functions. The cross sections of QCD subprocess are calculated
usually in the leading order (LO), as well as in the next to leading
order (NLO) \cite{1,2,NDE,Beer,Beer1}. The transverse momenta of the
incident partons are neglected in the QCD matrix elements. This is the
direct analogy of the Weizsaecker-Williams approximation in QED. It
allows to describe reasonably \cite{FMNR} the experimental data on
the total cross sections and one-particle distributions of produced
heavy flavours\footnote{New FNAL data \cite{Abb} on beauty production
are 2-3 times higher than the NLO parton model predictions.}, however
it can not reproduce, say, the azimuthal correlations \cite{MNR} of
two heavy quarks and the distributions of total transverse momentum
of heavy quarks pairs \cite{FMNR}, which are
determined by the transverse momenta of the incident partons.

There is an attempt to incorporate the transverse momenta of the incident
partons by a random shift of these momenta ($k_T$ kick) \cite{FMNR}
according to certain exponential distributions. This allows to describe
quantitatively the two-particle correlations \cite{FMNR}, but it
creates the problems in the simultaneous description of one-particle
longitudinal and transverse momentum distributions \cite{Shab}. At the
same time this procedure has no serious theoretical background. While
the shift of the order of $\Lambda_{QCD}$ ($\langle k_T \rangle \sim$
300 MeV) looks to be reasonable as having an possible origin in confinement
forces at large distances, the values $\langle k_T \rangle \sim$ 1 GeV
\cite{FMNR}, or even 3-4 GeV \cite{Apan,Nas} should be explained in
terms of the perturbative QCD.

Another possibility to incorporate the incident parton transverse
momenta is referred to as $k_T$-factorization approach
\cite{CCH,CE,MW,CH,CC,HKSST}, or the theory of semihard interactions
\cite{GLR,LR,8,lrs,3,SS,BS}.  Here the Feynman diagrams are
calculated taking account of the virtualities and of all possible
polarizations of the incident partons. In the small $x$ domain there
are no grounds to neglect the transverse momenta of the gluons,
$q_{1T}$ and $q_{2T}$, in comparison with the quark mass and
transverse momenta, $p_{iT}$.  Moreover, at very high energies and
very high $p_{iT}$ the main contribution to the cross sections comes
from the region of $q_{1T} \sim p_{1T}$ or $q_{2T} \sim p_{1T}$, see
Sect.~4 for details. The QCD matrix elements of the sub-processes are
rather complicated in such an approach. We have calculated them in
the LO. On the other hand, the multiple emission of soft gluons is
included here. That is why the question arises as to which approach
is more constructive.

The predictions of all phenomenological approaches rely
on the quark and gluon structure functions. The last ones are more or
less known experimentally from the data of HERA, but unknown at very
small values of Bjorken variable $x < 10^{-4}$. However it is just the
region that dominates in the heavy quark production at high
energies\footnote{For example, in the case of charm production, $m_c$ =
1.4GeV, at LHC, $\sqrt{s}$ = 14 TeV, the product $x_1x_2$ of two gluons
(both $x_1$ and $x_2$ are the integral variables) is equal to
$4\cdot 10^{-8}$ and applicapability of the existing structure
functions seems not to be clear at so small $x$, see discussion in \cite{ASS}.
Another problem at very small $x$ can be connected to gluon
shadowing \cite{LR,3,LRS1,ShT}}. A more serious
problem is probably related to the fact that the NLO parton model
formalism is based on the conventional structure functions
whereas the $k_T$-factorization
approach uses so-called unintegrated distributions which, at the
moment, are known with the accuracy not good enough.

In Sect. 2 we shortly discuss the conventional NLO parton model and
show some numerical results together with the experimental data.
The main formalism of the $k_T$-factorization approach is presented
in Sect.~3.

In Sect. 4 we present a comparison of results \cite{our} obtained with
the help of $k_T$-factorization and the parton model. The main goal of
this Section is to demonstrate the differences in the qualitative
predictions coming from the matrix elements. To simplify the
calculations and to avoid various additional dependences we have used a
"toy" gluon distribution which has only a reasonable qualitative
behaviour and a fixed value of $\alpha_S$.

More accurate comparison \cite{our1} between the
conventional parton model results and the $k_T$-factorization approach
is given in Sect.~5 for the experimentally measured quantities.
Here we have used more realistic gluon distribution GRV94 \cite{GRV} which is
compatible with the most recent data, see discussion in Ref.~
\cite{GRV1}. Predictions of the $k_T$-factorization approach for several
heavy quark correlations (some of them have never been discussed before)
are presented in Sect.~6.

\section{Conventional parton model approach}

The conventional parton model expression for the calculation of heavy
quark hadropro\-duc\-tion cross sections has the factorized form
\cite{CSS}:
\begin{equation}
\sigma (a b \rightarrow Q\overline{Q}) =
\sum_{ij} \int dx_i dx_j G_{a/i}(x_i,\mu_F) G_{b/j}(x_j,\mu_F)
\hat{\sigma} (i j \rightarrow Q \overline{Q}) \;,
\label{pm}
\end{equation}
where $G_{a/i}(x_i,\mu_F)$ and $G_{b/j}(x_j,\mu_F)$ are the structure
functions of partons $i$ and $j$ in the colliding hadrons $a$ and $b$,
$\mu_F$ is the factorization scale (i.e. the value of the order of
the maximal virtuality of incident partons) and $\hat{\sigma} (i j
\rightarrow Q \overline{Q})$ is the cross section of the subprocess
which is calculated in perturbative QCD. The latter cross section can
be written as a sum of LO and NLO contributions,
\begin{eqnarray}
\hat{\sigma} (i j \rightarrow Q\overline{Q}) &=&
\frac{\alpha_s^2(\mu_R)}{m_Q^2}\biggl(f^{(o)}_{ij}(\rho)
+ 4 \pi \alpha_s(\mu_R)
\Bigl[f_{ij}^{(1)}(\rho) + \bar f_{ij}^{(1)}(\rho)
\ln(\mu^2/m_Q^2)\Bigr]\biggr) \; ,
\label{shat}
\end{eqnarray}
where $\mu_R$ is the renormalization scale and $f^{(o)}_{ij}$ as
well as $f^{(1)}_{ij}$ and $\bar{f}^{(1)}_{ij}$ depend only on the single
variable
\begin{equation}
\rho = \frac{4m_Q^2}{\hat{s}} \;, \; \hat{s} = x_i x_j s_{ab} \;.
\end{equation}
(In the factor $\ln(\mu^2/m_Q^2)$ we assume $\mu_R = \mu_F$ following
\cite{1}. In the case of different values of $\mu_R$ and $\mu_F$, which
is preferable for the description of the experimental data \cite{FMNR},
Eq.~(2) becomes more complicated.)

The expression (1) corresponds to the process shown schematically in
Fig.~1 with $q_{iT} = q_{jT} =0$. The main contribution to the cross
section at small $x$ is known to come from gluons, $i = j = g$.

\begin{figure}[htb]
\begin{center}
\mbox{\epsfig{file=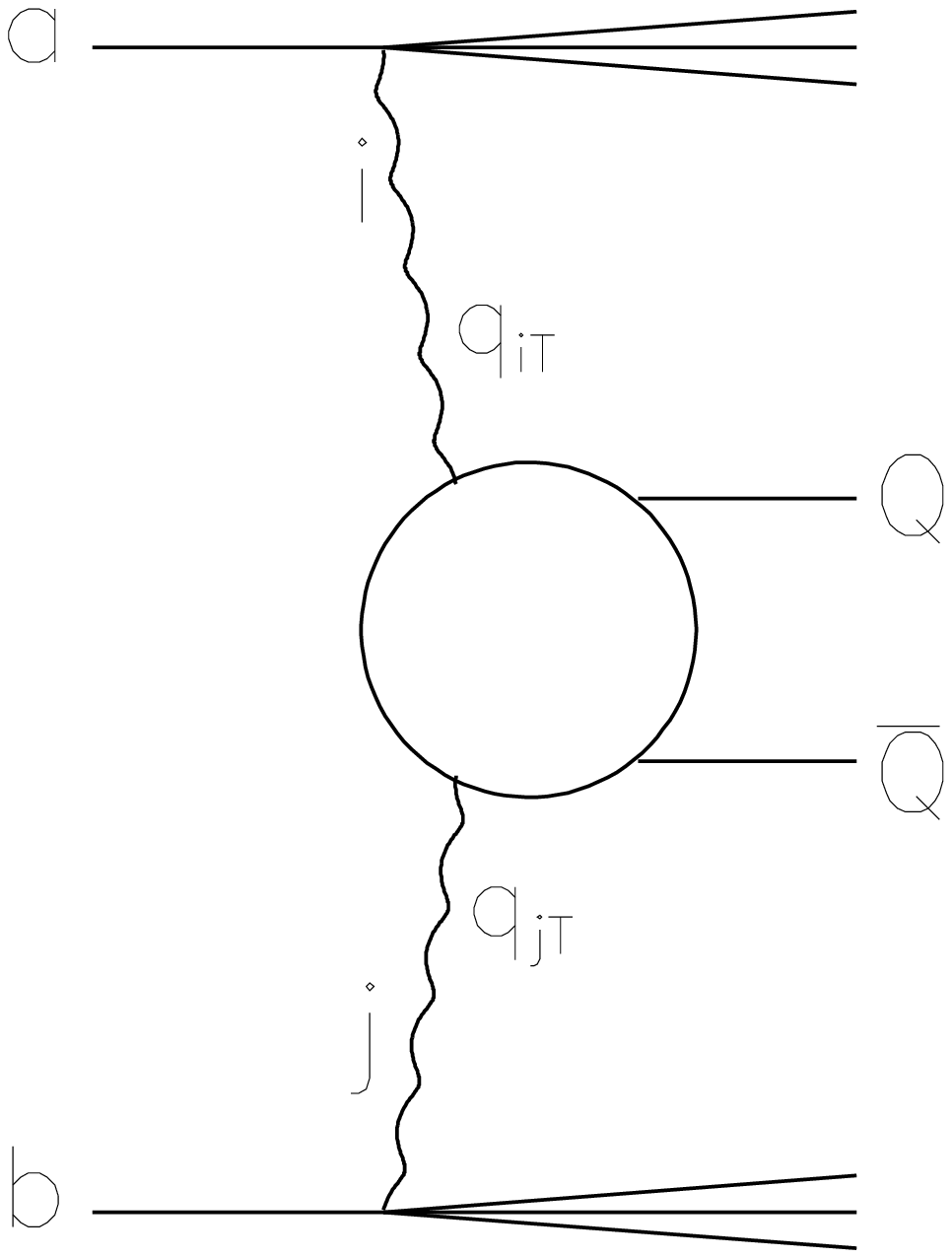,width=0.60\textwidth}} \\
Fig. 1. Heavy quark production in hadron-hadron collision. The LO parton
model corresponds to the case when $q_{1t}=q_{2t}=0$.
\end{center}
\end{figure}

The principal uncertainties of any numerical QCD calculation are
the consequences of unknown scales \footnote{These uncertainties have
to disappear after all high order contributions are summed up. There
is the opinion that the strong (weak) scale dependence in LO or NLO
means the large (small) contribution of high order diagrams, but it
is not the case. The strong or weak scale dependence in LO or NLO
indicates only the same dependence for the higher orders,
which could
be numerically small or large at some particular fixed scale.}, $\mu_F$
and $\mu_R$ and the heavy quark mass, $m_Q$. Both scales (sometimes they
are assumed to be equal) are to be of the order of hardness of the
treated process, however which value is better to take,
$m_Q$, $m_T = \sqrt{m_Q^2 + p_T^2}$ or $\hat{s}$, remains to be
unknown. In principle, the uncertainties should not be essential
because of the logarithmical dependence on these parameters.
Unfortunately the masses of $c$ and $b$ quarks are not large enough
and it leads to numerically large uncertainties (see e.g. \cite{FMNR})
at the existing energies for fixed target experiments.

Usually in the parton model the values
\begin{equation}
\mu_F = \mu_R = m_Q
\end{equation}
are used for the total cross sections and
\begin{equation}
\mu_F = \mu_R = m_T = \sqrt{m_Q^2 + p_Q^2}
\end{equation}
for the one-particle distributions \cite{FMNR}.

Here we calculate the total cross sections of heavy quark production
as the integrals over their $p_T$ distrubutions, i.e. with the scales~(5).

Both in the parton model and in the $k_T$-factorization approach we take
\begin{equation}
m_c = 1.4\; {\rm GeV}, \qquad m_b = 4.6\; {\rm GeV} \;,
\end{equation}
for the values of short-distance perturbative quark masses
\cite{Nar,BBB}.

Another principal problem of the parton model is the collinear
approximation. The transverse momenta of the incident partons, $q_{iT}$
and $q_{jT}$ are assumed to be zero, and their virtualities are
accounted for through the structure functions only; the cross section
$\hat{\sigma} (i j \rightarrow Q \overline{Q})$ is assumed to be
independent of these virtualities. Naturally, this approximation
significantly simplifies the calculations.

The NLO parton model calculations of the total cross sections of
$c\bar{c}$ and $b\bar{b}$ production, as functions of the beam
energy, for $\pi^- N$ and $p-N$ collisions can be found in \cite{FMNR}.
These results depend strongly (by the factor of several times) on the
numerical values of quark masses as well as on the both scales, $\mu_F$
and $\mu_R$. Although there is a contradiction in some experimental data
generally (at fixed target energy) they are in
agreement with NLO parton model predictions.

The NLO corrections to one-particle distributions lead only to
renormalization of the LO results, practically without changing in the
shapes of the distributions \cite{NDE1,MNR}. It means that instead of the
more complicated NLO calculation of the $p_T$ or rapidity distributions
it is enough to calculate them in LO, and to multiply then by K-factor
\begin{equation}
K = \frac{LO + NLO}{LO} \;,
\end{equation}
which can be taken, say, from the results on the total production cross
sections. The comparison of LO + NLO calculations with the LO multiplied by
the K-factor is presented in Fig. 2 taken from Ref. \cite{NDE1}.

\begin{figure}[htb]
\begin{center}
\mbox{\epsfig{file=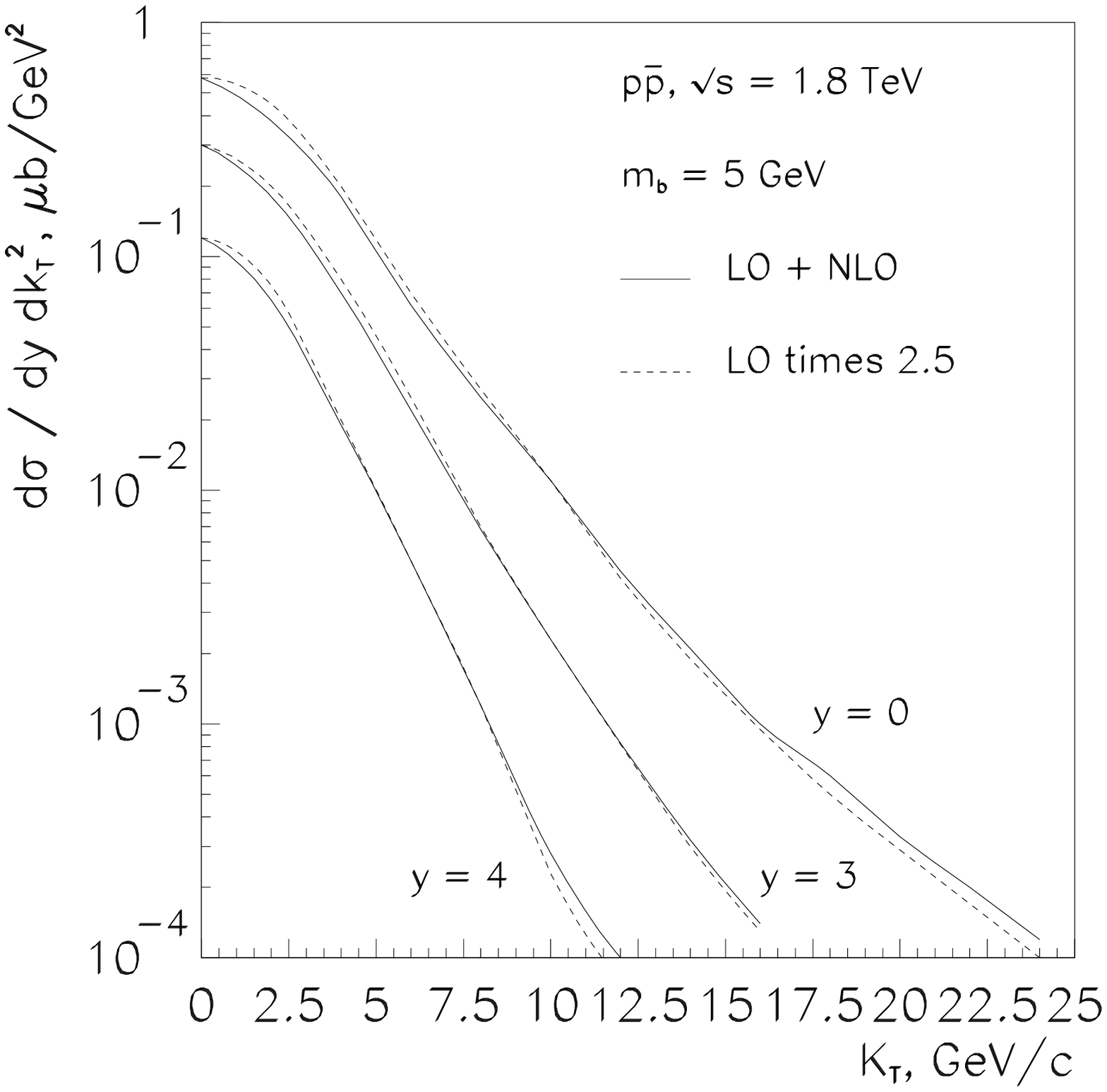,width=0.60\textwidth}} \\
Fig. 2. The calculated $p_T$-distributions for $p\bar{p} \rightarrow
b + X$ at $\sqrt{s}$ = 1.8 TeV, for different values of rapidity
for LO+NLO QCD parton model and for LO contribution multiplied by
K-factor, Eq.~(7) equal to 2.5.
\end{center}
\end{figure}

The values of K-factors and their energy and scale dependences for
the various sets of structure functions were calculated in Refs.
\cite{SCG,LSCSh}. These dependences are similar for the modern sets
of structure functions.

The experimental data of WA92 \cite{Adam} and E769 \cite{Alv}
for $x_F$-distributions of D-mesons produced in $\pi N$ interactions
are in agreement with the parton model distributions for bare quarks, as
one can see in Fig.~3 taken from \cite{FMNR}. It means that the
fragmentation processes are not important here, or they are compensated
by, say, recombination processes. The shape of $x_F$-distributions does
not depend practically on the mass of $c$-quark.

\begin{figure}
\begin{center}
\centerline{\epsfig{figure=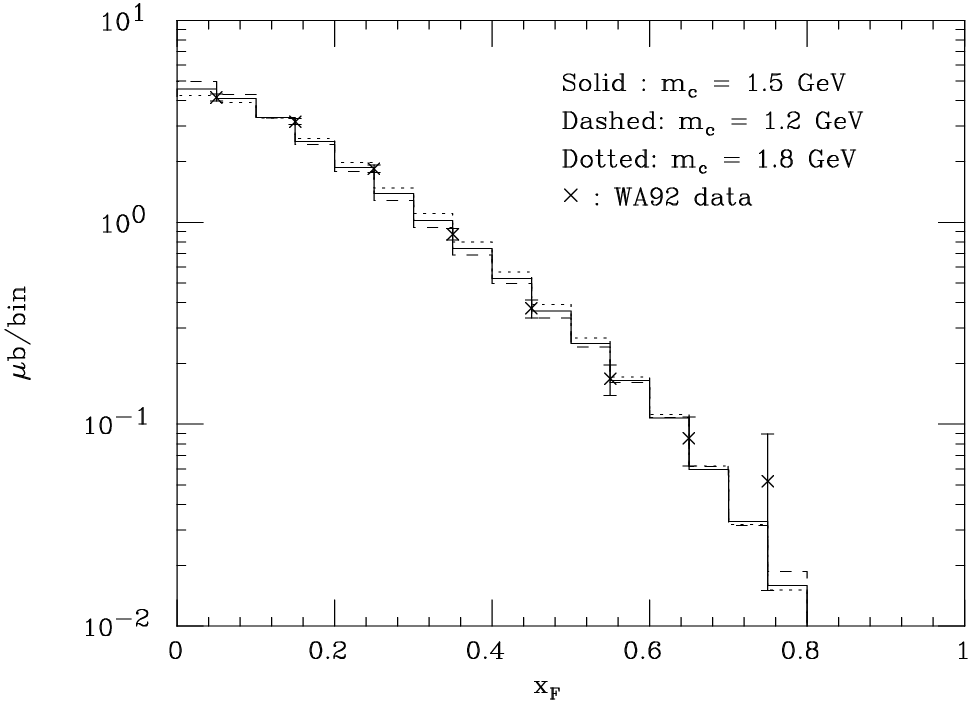,width=0.50\textwidth,clip=}
            \hspace{0.3cm}
            \epsfig{figure=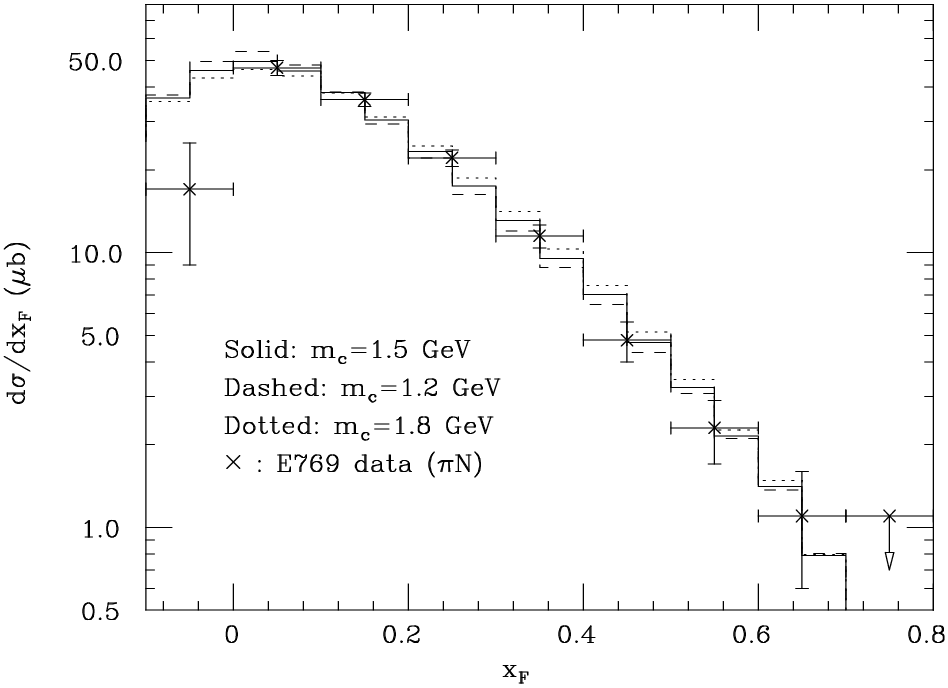,width=0.50\textwidth,clip=}}
Fig. 3. Experimental $x_F$ distributions for $D$ mesons, compared to the
NLO parton model prediction for charm quarks.
\end{center}
\end{figure}

It was claimed in \cite{FMNR} that the data on one-particle
$p_T$-distributions, both obtained by fixed target experiments and by
hadronic colliders (D0, CDF and UA1 collaborations) for the case of
beauty production can be described by the NLO parton model using the
variation of both scales, $\mu_R$ and $\mu_F$. However new FNAL data
on beauty production \cite{Abb} are 2-3 times higher than the
theoretical curve \cite{MNR}.

More serious problem appears when we consider the correlations of
produced heavy quarks. Let as start from so-called azimuthal
correlation. The azimuthal angle $\phi$ is defined as an opening angle
between two produced heavy quarks, projected onto a plane perpendicular
to the beam and defined as the $xy$-plane. In the LO parton model the
sum of the produced heavy quark momenta projected onto this plane is
exactly zero, and the angle between them is always 180$^o$. In the
case of the NLO parton model the $\phi$ distribution is non-trivial
\cite{MNR}, however the predicted distribution (without including the
$k_T$ kick) is narrower in comparison with the fixed target data
\cite{FMNR}.

The investigation of such distributions is very important. The
relative LO and NLO contributions of the parton model depend on the
unknown scales, $\mu_F$ and $\mu_R$. In the one-particle
distributions, the sum of these contributions practically coinsides
with the LO contribution multiplied by the $K$-factor, as it is shown in
Fig.~2.  Therefore we are unable to control separately the magnitudes of
the LO and NLO contributions. Contrary, in the case of azimuthal
correlations all difference from the simple $\delta(\phi - \pi)$
distribution comes from the NLO contribution.

The experimental data on azimuthal correlations are claimed (see
\cite{BEAT,BEAT1} and Refs. therein) to be in disagreement with the NLO
predictions, for the cases of charm pair hadro- and photoproduction at
fixed target energies. The level of the disagreement can be seen in Fig.~4
(solid histograms) taken from Ref.~\cite{FMNR}. These data can be
described \cite{FMNR}, assuming the comparatively large intrinsic
transverse momenta of incoming partons ($k_T$ kick). For each
event, in the longitudinal centre-of-mass frame of the heavy quark
pair, the $Q\overline{Q}$ system is boosted to rest. Then a second
transverse boost is performed, which gives the pair a transverse
momentum equal to $\vec{p}_T(Q\overline{Q}) = \vec{k}_T(1) +
\vec{k}_T(2)$; $\vec{k}_T(1)$ and $\vec{k}_T(2)$ are the transverse
momenta of the incoming partons, which are chosen randomly, with
their moduli distributed according to
\begin{equation}
\frac{1}{N}\frac{dN}{dk_T^2}=\frac{1}{\langle k_T^2 \rangle}
      \exp(-k_T^2/\langle k_T^2 \rangle).
\end{equation}

\begin{figure}
\begin{center}
\centerline{\epsfig{figure=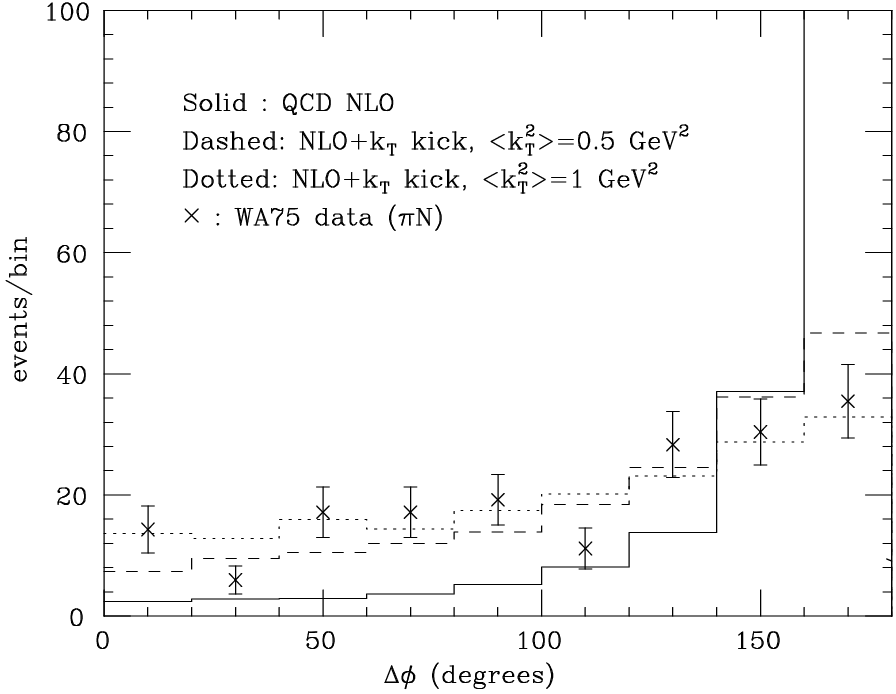,width=0.50\textwidth,clip=}
            \hspace{0.3cm}
            \epsfig{figure=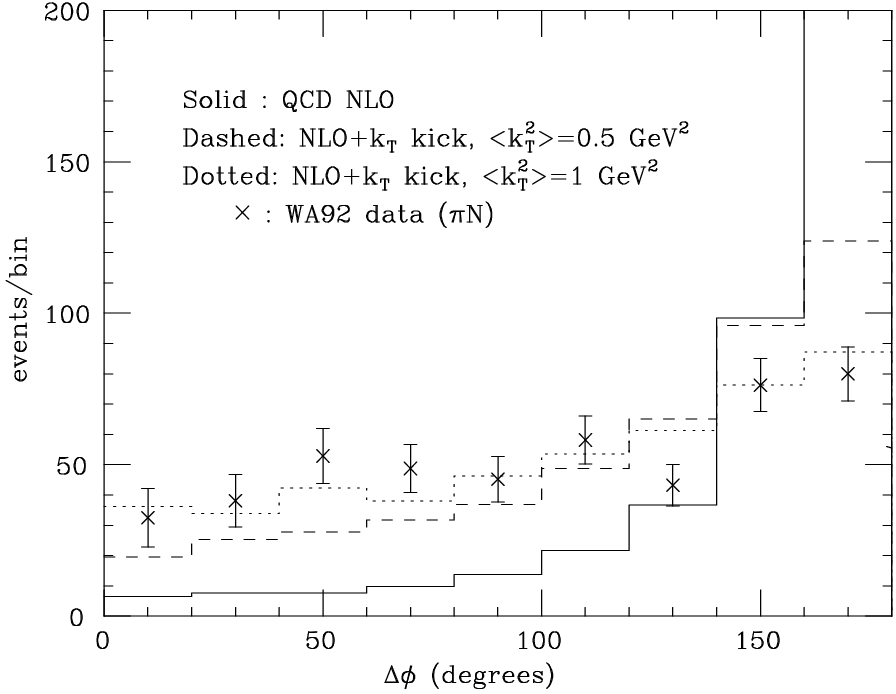,width=0.50\textwidth,clip=}}
Fig. 4. Azimuthal correlation for charm production in $\pi N$
collisions: NLO parton model and $k_T$ kick calculations versus the WA75
and WA92 data.
\end{center}
\end{figure}

The dashed and dotted histograms in Fig. 4 correspond to the NLO
parton model prediction, supplemented with the $k_T$ kick with
$\langle k_T^2\rangle=0.5$ GeV$^2$ and $\langle k_T^2\rangle=1$ GeV$^2$,
respectively. We see that with $\langle k_T^2 \rangle= 1$ GeV$^2$ it
is possible \cite{FMNR} to describe the data.

However the large intrinsic transverse momentum significantly changes
one-particle $p_T$-distributions of heavy flavour hadrons, which were
earlier in good agreement with the data. The solid curves in Fig.~5
taken from \cite{FMNR} represent the NLO parton model predictions for
charm quarks $p_T$-distributions which are in agreement with the data.
The effect of the $k_T$ kick results in a hardening of the $p_T^2$
spectrum. On the other hand, by combining the $k_T$ kick with
$\langle k_T^2 \rangle =1$ GeV$^2$ and the Peterson fragmentation
\cite{Pet}, the theoretical predictions slightly undershoot the data
(dot-dashed curves). For the predictions at higher energy the values
$\langle k_T^2 \rangle =$ 3-4 GeV$^2$ \cite{Apan,Nas} were used
without any connection with perturbative QCD diagrams.

\begin{figure}
\begin{center}
\centerline{\epsfig{figure=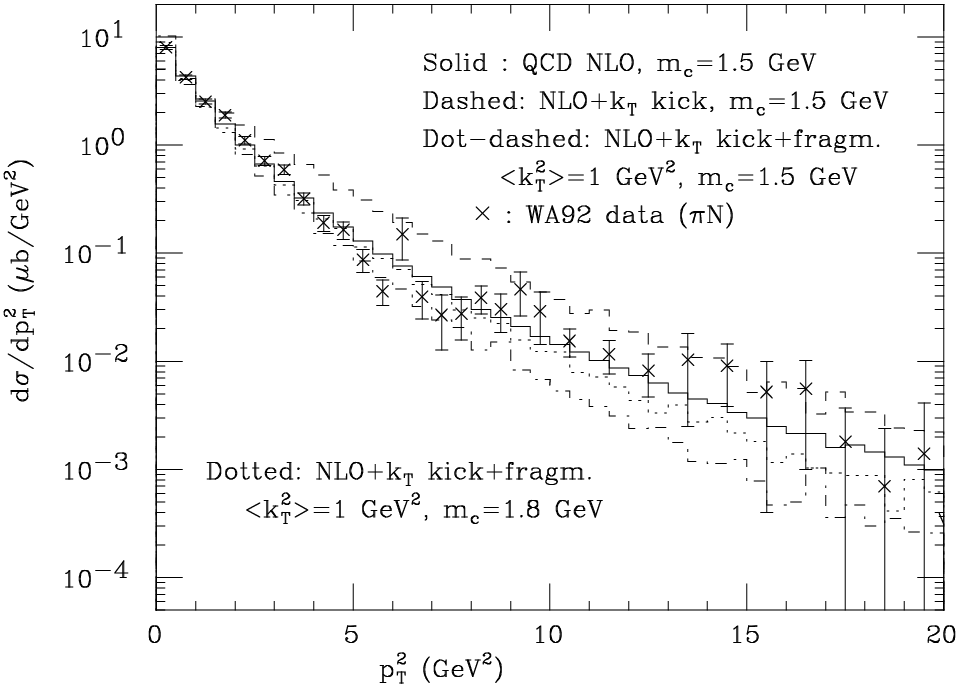,width=0.50\textwidth,clip=}
            \hspace{0.3cm}
            \epsfig{figure=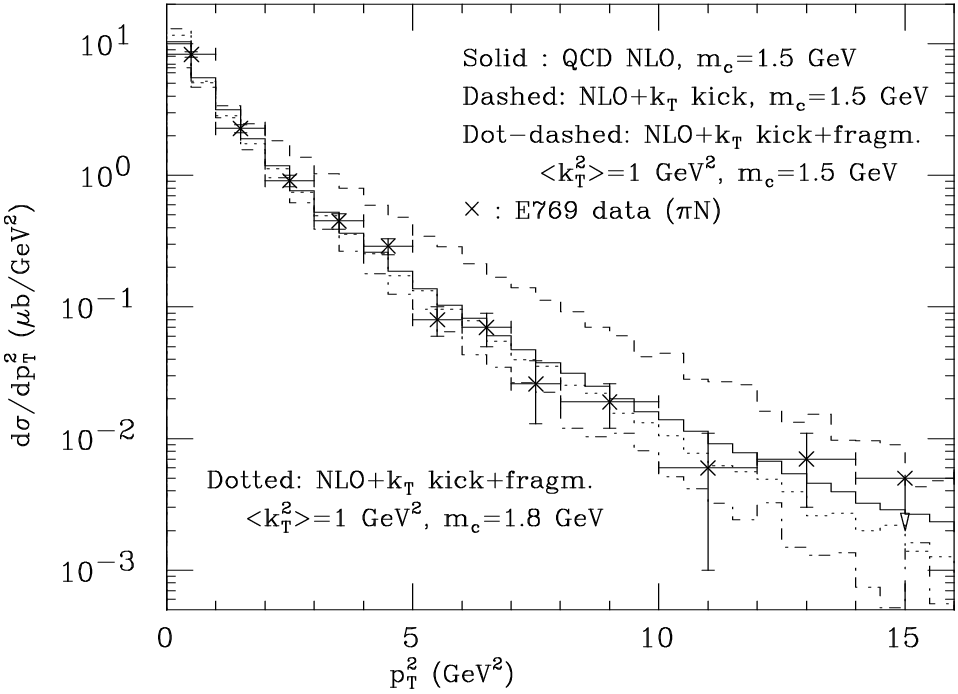,width=0.50\textwidth,clip=}}
Fig. 5. Charm $p_T^2$ distribution measured by WA92 and E769, compared
to the NLO parton model predictions, with and without the
non-perturbative effects.
\end{center}
\end{figure}

The $k_T$ kick affects the $x_F$-distributions of the
produced $c$-quarks very weakly, Fig.~3, and after accounting
the fragmentation these distributions become too soft.

The general phenomenological expression for the heavy quark
production can be written\footnote{We omit here for the simplicity all
unimportant factors.} as a convolution of the initial
transverse momenta distributions, $I(q_{1T})$ and $I(q_{2T})$,
with squared modulo of the matrix element,
\begin{equation} \sigma_{QCD}(Q\overline{Q})\ \propto\
\int d^2 q_{1T} d^2 q_{2T} I(q_{1T}) I(q_{2T}) \vert M(q_{1T},
q_{2T}, p_{1T}, p_{2T}) \vert ^2 \;.
\label{ge}
\end{equation}

Here there are two possibilities: \\
i) the typical gluon transverse momenta are much
smaller than the transverse momenta of produced heavy quarks,
$q_{iT} \ll p_{iT}$, or \\
ii) all transverse momenta are of the same order,
$q_{iT} \sim p_{iT}$.

In the first case one can replace both initial distributions
$I(q_{iT})$ by $\delta$-functions. It reduces the expression (\ref{ge})
to the simple form (collinear approximation):
\begin{equation}
\sigma_{coll.}(Q\overline{Q})\ \propto\ \vert
M(0, 0, p_{1T}, p_{2T}) \vert ^2
\end{equation}
in total agreement with Weizsaecker-Williams approximation.

In the second case we can not a priory expect good results from the
Weizsaecker-Williams approximation. However it gives the reasonable
numerical results in some cases.

The $k_T$ kick \cite{FMNR} discussed above effectively accounts for the
transverse momenta of incident partons. It is based on the expression which
symbollically reads
\begin{equation}
\sigma_{kick}(Q\overline{Q})\ \propto\ I(q_{1T}) I(q_{2T})
\otimes \vert M(0, 0, p_{1T}, p_{2T}) \vert ^2\
\end{equation}
that has no good theoretical background.

The main difference from the general QCD expression, Eq.~(\ref{ge}),
is that due to the absence of $q_{iT}$ in the matrix element the value
$\langle k_T^2\rangle$ has to be differently taken for different
processes (say, for heavy flavour production with comparatively small
or large $p_T$). The reason is that the functions $I(q_{iT})$ in general
QCD expression decrease for large $q^2_{iT}$ as a weak power (see next
Sect.), i.e. comparatively slowly, and the $q^2_{iT}$ dependence of
the matrix element is more important.

In the last one the corrections of the order of $q^2_{iT}/\mu^2$,
where $\mu^2$ is the QCD scale, are small enough when
$q^2_{iT}/\mu^2 << 1$ and they start to suppress the matrix element
value when $q^2_{iT}/\mu^2 \sim 1$. Thus the essential values of the
$q^2_{iT}$ depend on the hardness of the process.

\section{Heavy quark production in the $k_T$-factorization approach}

The transverse momenta of the incident gluons in the small-$x$ region
result in the $k_T$-factorization approach from
$\alpha_S \ln k_T^2$ gluon diffusion. The diffusion is
described by the so-called unintegrating gluon distribution - the
function $\varphi(x,q^2)$ giving the gluon density at the fixed
fraction of the longitudinal momentum of the initial hadron, $x$, and
the gluon virtuality, $q^2$. These distributions should be found
with the help of evolution equation and the experimental data.
Unfortunately, it is not yet fulfilled. At very low $x$, that is to
leading $\log(1/x)$ accuracy, the unintegrating gluon distribution
are approximately determined \cite{GLR} via the derivative of the usual
structure function:
\begin{equation}
\label{xG}
\varphi (x,q^2)\ =\ 4\sqrt2\,\pi^3
\frac{\partial[xG(x,q^2)]} {\partial q^2}\ .
\end{equation}
Such a definition of $\varphi(x,q^2)$ enables to treat correctly the
effects arising from the gluon virtualities.

Although generally $\varphi$ is a function of three variables, $x$,
$q_T$ and $q^2$, the transverse momentum dependence is comparatively
weak since $q_T^2 \approx - q^2$ for small $x$ in LLA in agreement with
$q^2$-dependences of structure function. Note that due to QCD scaling
violation at larger $q^2$, the value of $\varphi(x,q^2)$ for the
realistic structure functions increases faster with decreasing of
$x$. Therefore the larger $q_T$ becomes important at small $x$ in the
numerical calculations.

The exact expression of the $q_T$ gluon distribution can be obtained as a
solution of the evolution equation which, contrary to the parton model
case, is nonlinear due to interactions between the partons in the small
$x$ region. The calculations \cite{Blu} of the $q_T$ gluon distribution
in leading order using the BFKL theory \cite{BFKL} result in differences
from our $\varphi(x,q^2)$ function given by Eq.~(12) by only about
10-15\%.

Here we deal with the matrix element accounting for the gluon
virtualities and polarizations. Since it is much more complicated than
in the parton model we consider only the LO of the subprocess
$gg \to Q\bar{Q}$ which gives the main contribution to the heavy quark
production cross section at small $x$, see the diagrams a, b and c
in Fig.~6.  The lower and upper ladder blocks present the
two-dimensional gluon functions $\varphi(x_1,q_1^2)$ and
$\varphi(x_2,q_2^2)$.

\begin{figure}
\begin{center}
\mbox{\epsfig{file=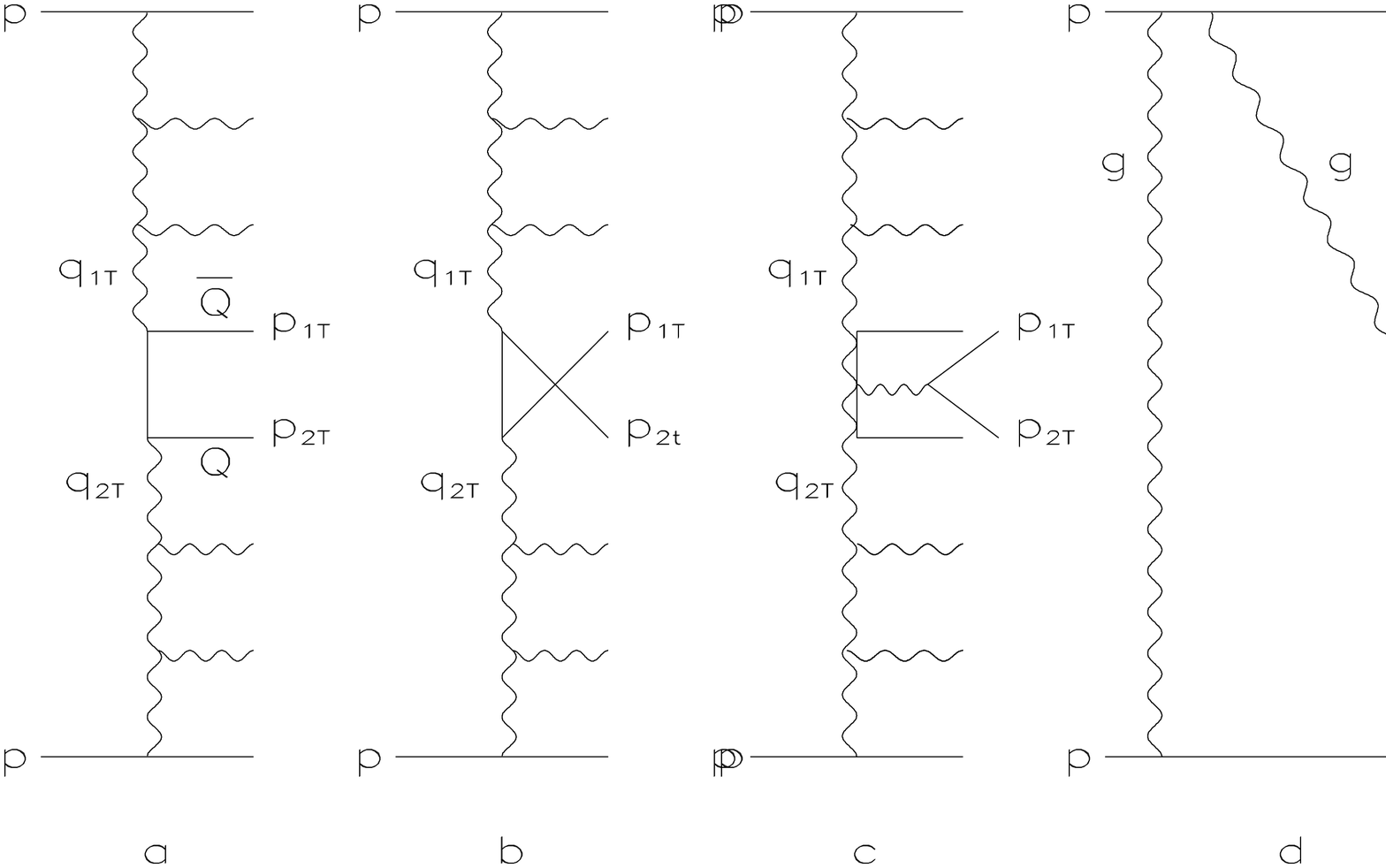,width=0.60\textwidth}} \\
Fig. 6. Low order QCD diagrams for heavy quark production in $pp$
($p\overline{p}$) collisions via gluon-gluon fusion (a-c) and the
diagram (d) formally violating the factorization and being necessary
for gaude invariance.
\end{center}
\end{figure}

Strictly speaking, Eq. (12) may be justified in the leading $\log (1/x)$
limit only. The unintegrated parton distribution
$f_a(x,q_T,\mu)$ determines the probability to find a parton $a$ initiating
the hard process with the transverse momentum $q_T$ (and with
factorization scale $\mu$). To restore the function $f_a(x,q_T,\mu)$ on
the basis of the conventional
(integrated) parton
density $a(x,\lambda^2)$ we have to consider the DGLAP
evolution\footnote{For the $g \to gg$ splitting we need to insert a
factor $z'$ in the last integral of Eq. (13) to account for the
identity of the produced gluons.}
\begin{equation}
\frac{\partial a}{\partial \ln \lambda^2} = \frac{\alpha_s}{2 \pi}
\left[ \int^{1-\Delta}_x P_{aa'} (z) a'(\frac xz,\lambda^2) dz -
a(x,\lambda^2) \sum_{a'} \int^{1-\Delta}_0 P_{a'a} (z') dz' \right]
\end{equation}
(here $a(x,\lambda^2)$ denotes $xg(x,\lambda^2)$ or $xq(x,\lambda^2)$
and $P_{aa'}(z)$ are the splitting functions).

The first term on the right-hand side of Eq. (13) describes the number
of partons $\delta_a$ emitted in the interval
$\lambda^2 < q^2_T < \lambda^2+\delta \lambda^2$, while the second
(virtual) term reflects the fact that the parton $a$ disappears after the
splitting.

The second contribution may be resumed as giving survival probability
$T_a$ that the parton $a$ with the transverse momentum $q_T$ remains
untouched in the evolution up to the factorization scale
\begin{equation}
\label{Sud}
T_a(q_T,\mu) = \exp \left[ -\int^{\mu^2}_{q^2_T}
\frac{\alpha_s(p_T)}{2\pi} \frac{dp^2_T}{p^2_T}
\sum_{a'} \int^{1-\Delta}_0 P_{a'a} (z') dz' \right]
\end{equation}

For the case of one loop QCD running coupling
$\alpha_s(p_T) = 4\pi /b \ln{p^2_T/\Lambda^2}$ the factor
$T_a(q_T,\mu)$ can be written down explicitely. In particular, the gluon
survival probability (which enters our formulae) reads:
\begin{eqnarray}
& T_a ( q_T, \mu ) = \frac{\ln(\mu /\Lambda)}{\ln(q_T /\Lambda)}
\cdot \exp \left\{ \frac{8N_c}{b} \left[ \ln
\frac{\mu}{q_T} - \ln \frac{\mu}{\Lambda} \ln
\frac{\ln(\mu /\Lambda )}{\ln(q_T/\Lambda)} - 2E_1(q_T,\mu) +
3/2 E_2(q_T,\mu) - \right. \right. \nonumber \\
\nonumber \\
& \left. -  2/3 E_3(q_T,\mu) + 1/4 E_4(q_T,\mu) \right]
+ \\
\nonumber \\
&\left. + \frac{2}{3b}n_F \left[ 3E_1(q_T,\mu) - 3E_2(q_T,\mu) +
2E_3(q_T,\mu) \right] \right\} \nonumber
\end{eqnarray}
where
\begin{equation}
E_k(q_T,\mu) = \left( \frac{\Lambda}{\mu} \right)^k
[Ei(k\ln(\mu/\Lambda ) - Ei(k\ln(q_T/\Lambda )) ] \;,
\end{equation}
and the integral exponent $Ei(z) = \int^z_{-\infty}\frac{dt}{t}\exp{t}$;
$n_F$ and $N_c$ are the number of light quark flavoures and the
number of colours, respectively, and $b = 11 - \frac23 n_F$.

Thus the unintegrated distribution $f_a(x,q_T,\mu)$ reads
\begin{equation}
f_a(x,q_T,\mu) = \left[\frac{\alpha_s}{2 \pi}
%\Theta (1-\Delta-x)
\int^{1-\Delta}_x P_{aa'} (z) a'\left(\frac xz,q_T^2 \right) dz
\right] T_a(q_T,\mu) \;,
\end{equation}
where the cut-off $\Delta = q_T/\mu$ is used both in Eqs. (14) and
(17) \cite{MW1,KMR}.

We have to emphasize that $f_a(x,q_T,\mu)$ is just the quantity which
enters into the Feynman diagrams. The distributions
$f_a(x,q_T,\mu)$ involve two hard scales\footnote{This property is
hidden in the conventional parton distributions as $q_T$ is
integrated up to the scale $\mu$.}: $q_T$ and the scale $\mu$ of the
probe. The scale $\mu$ plays a dual role. On the one hand side it acts as
the factorization scale, while on the other hand side it controls the
angular ordering of the partons emitted in the evolution
\cite{MCi,CFM,Mar}. The factorization scale $\mu$ separates the
partons associated with the emission off
both the beam and target protons (in $pp$ collisions) and
off the hard subprocess. For example, it separates emissions off
the beam (with polar angle $\theta < 90^o$ in c.s.m.) from those off
the target (with $\theta > 90^o$ in c.s.m.), and from the
intermediate partons from the hard subprocess. This separation was
proved in \cite{MCi,CFM,Mar} and originates from the destructive
interference of the different emission amplitudes (Feynman diagrams)
in the angular boundary regions.

If the longitudinal momentum fraction is fixed by the hard
subprocess, then the limits of the angles can be expresseed in terms
of the factorization scale $\mu$ which corresponds to the upper limit
of the allowed value of the $s$-channel parton $k_T$.

Since the parton distributions contain two scales, one has to deal
with a complicated evolution equation for $f_a(x, q_T, \mu)$. On the
other hand since the evolution is controlled by the emission angle
only, the distribution $f_a(x, q_T, \mu)$ can be obtained from a
single scale evolution equation, as it was done in Eq.~(17).

In the leading $\log (1/x)$ (i.e. BFKL) limit the virtual DGLAP
contribution is neglected. So $T_a = 1$ and one comes back to Eq.~(12)
\begin{equation}
f_a^{BFKL} (x,q_T,\mu) = \frac{\partial a(x,\lambda^2)}
{\partial \ln \lambda^2}\;\;\;, \lambda = q_T \; .
\end{equation}

In the double log limit Eq. (17) can be written in the form
\begin{equation}
f_a^{DDT}(x,q_T,\mu) = \frac\partial{\partial \ln \lambda^2}
\left[a(x,\lambda^2) T_a(\lambda,\mu)\right]_{\lambda = q_T} \; ,
\end{equation}
which was firstly proposed by \cite{DDT}. In this limit the
derivative $\partial T_a /\partial \ln \lambda^2$ cancels the second
term of the r.h.s. of Eq. (13) (see \cite{KMR} for a more detailed
discussion).

Finally, the probability $f_a(x,q_T,\mu)$ is related to the BFKL
function $\varphi(x,q^2_T)$ by
\begin{equation}
\varphi (x,q^2)\ =\ 4\sqrt2\,\pi^3 f_a(x,q_T,\mu) \; .
\end{equation}

Note that due to a virtual DGLAP contribution the derivative
$\partial a(x,\lambda^2) / \partial \lambda^2$ can be negative for
not small enough $x$ values. This shortcoming of Eq.~(18) is partly
overcome in the case of Eq.~(19). Unfortunately the cut-off $\Delta$
used in a conventional DGLAP computation does not depend on the scale
$\mu$. To obtain an integrated parton distributions it is enough to put
any small $\Delta \ll 1$\footnote{There is a cancellation between the
real and virtual soft gluon DL contributions in the DGLAP equation, written
for the integrated partons (including all $k_T\le\mu$). The emission of a
soft gluon with momentum fraction $(1-z)\to 0$ does not affect the
$x$-distribution of parent partons. Thus the virtual and real
contributions originated from $1/(1-z)$ singularity of the splitting
function $P(z)$ cancel each other.

On the contrary, in the unintegrated case the emission of soft gluon
(with $q'_T>k_T)$ alters the transverse momentum of parent ($t$-channel)
parton. Eq.~(17) includes this effect through the derivative
$\partial T(k^2_T,\mu^2)/\partial k^2_T$.}.

On the other hand, in the survival probability Eq. (14) we have to
use the true (within the leading log approximation) value $\Delta =
q_T/\mu$.  Hence for a rather large $q_T$ ($\sim \mu$) and
$x$ even the DDT form Eq.~(19) is not precise. Only the
expression (17) with the same cut-off $\Delta$ in a real DGLAP
contribution and in the survival probability (14) provides the positivity
of the unintegrated probability $f_a(x,q_T,\mu)$ in the whole
interval $0 < x < 1$.

Of course, just by definition $f_a(x,q_T,\mu) = 0$ when the
transverse momentum $q_T$ becomes larger than the factorization scale
$\mu$.

In what follows we use the Sudakov decomposition for the quark momenta
$p_{1,2}$ through the momenta of colliding hadrons  $p_A$ and
$p_B\,\, (p^2_A = p^2_B \simeq 0)$ and the transverse momenta
$p_{1,2T}$:
\begin{equation}
\label{1}
p_{1,2} = x_{1,2} p_B + y_{1,2} p_A + p_{1,2T}.
\end{equation}
The differential cross section of heavy quark hadroproduction has the
following form:\footnote{We put the argument of $\alpha_S$ to be equal
to gluon
virtuality, which is very close to the BLM scheme\cite{blm}; (see also
\cite{lrs}).}
\begin{eqnarray}
\frac{d\sigma_{pp}}{dy^*_1 dy^*_2 d^2 p_{1T}d^2
p_{2T}}\,&=&\,\frac{1}{(2\pi)^8}
\frac{1}{(s)^2}\int\,d^2 q_{1T} d^2 q_{2T} \delta (q_{1T} +
q_{2T} - p_{1T} - p_{2T}) \nonumber \\
\label{spp}
&\times &\,\frac{\alpha_s(q^2_1)}{q_1^2} \frac{\alpha_s (q^2_2)}{q^2_2}
\varphi(q^2_1,y)\varphi (q^2_2, x)\vert M_{QQ}\vert^2.
\end{eqnarray}
Here $s = 2p_A p_B\,\,$, $q_{1,2T}$ are the gluon transverse momenta
and $y^*_{1,2}$  are the quark rapidities in the hadron-hadron c.m.s.
frame,
\begin{equation}
\label{xy}
\begin{array}{lcl}
x_1=\,\frac{m_{1T}}{\sqrt{s}}\, e^{-y^*_1}, &
x_2=\,\frac{m_{2T}}{\sqrt{s}}\, e^{-y^*_2},  &  x=x_1 + x_2\\
y_1=\, \frac{m_{1T}}{\sqrt{s}}\, e^{y^*_1}, &  y_2 =
\frac{m_{2T}}{\sqrt{s}}\, e^{y^*_2},  &  y=y_1 + y_2 \\
&m_{1,2T}^2 = m_Q^2 + p_{1,2T}^2. &
\end{array}
\end{equation}
$\vert M_{QQ}\vert^2$ is the square of the matrix element for the heavy
quark pair hadropro\-duction. Contrary to the mention of \cite{BS},
the transformation Jacobian from $x, y$ to $y^*_1, y^*_2$ is
accounted in our matrix element.

In LLA kinematic
\begin{equation}
\label{q1q2}
\begin{array}{crl}
q_1 \simeq \,yp_A + q_{1T}, & q_2 \simeq \,xp_B + q_{2T}.
\end{array}
\end{equation}
so
\begin{equation}
\label{qt}
\begin{array}{crl}
q_1^2 \simeq \,- q_{1T}^2, & q_2^2 \simeq \,- q_{2T}^2.
\end{array}
\end{equation}
(The more accurate relations are $q_1^2 =- \frac{q_{1T}^2}{1-y}$,
$q_2^2 =- \frac{q_{2T}^2}{1-x}$ but we are working in the kinematics
where $x,y \ll 1$).

The matrix element $M$ is calculated in the Born approximation of QCD
without standard simplifications of the parton model.

In the axial gauge
$p^\mu_B A_\mu = 0$ the gluon propagator takes the form
$D_{\mu\nu} (q) = d_{\mu\nu} (q)/q^2,$
\begin{equation}
\label{prop}
d_{\mu\nu}(q)\, =\, \delta_{\mu\nu} -\, (q^\mu p^\nu_B \, + \, q^\nu
p^\mu_B )/(p_B q) \;.
\end{equation}

For the gluons in $t-$channel the
main contribution comes from the so called "nonsense" polarization
$g^n_{\mu\nu}$, which can be picked out by decomposing the numerator
into longitudinal and transverse parts:
\begin{equation}
\label{trans}
\delta_{\mu\nu} (q)\ =\ 2(p^\mu_B p^\nu_A +\, p^\mu_A p^\nu_B )/s\,
+ \, \delta^T_{\mu\nu} \approx\, 2p^\mu_B p^\nu_A /s\,\equiv\,
g^n_{\mu\nu}\ .
\end{equation}
The other contributions are suppressed by the powers of $s$. It is easy
to check that in axial gauge (\ref{prop}) $p_B^\mu d_{\mu \nu}=0$ and
$p_A^\mu d_{\mu \nu} \simeq -q_T^ \nu/y$. Thus we effectively get
\begin{equation}
\label{trans1}
d_{\mu\nu} (q)\ \approx\ -\,2\, \frac{p^\mu_B q^\nu_T}{y\,s}\ .
\end{equation}
Another way to obtain the same result is to use the transversality
condition. Since the sum of the diagram Fig.~6a-c is gauge invariant
(see below) the product $q_{1\mu} M_\mu = 0$, here $M_\mu$ denotes
the amplitude of the gluon $q_1$ and hadron $p_B$ interaction
described by the lower part of Fig.~6a--c.  Using the expansion
(\ref{q1q2}) for the $q_1$ momentum we obtain $$ p_A^\mu M_\mu\
\simeq\ -\frac{q_{1T}^\mu}y M_\mu\ , $$ which leads to
Eq.~(\ref{trans1}) or
\begin{equation}
\label{trans2}
d_{\mu\nu} (q)\
\approx\ \,2\, \frac{q^\mu_T q^\nu_T}{xys}\ ,
\end{equation}
if we do such a trick for the vector $p_B$ too \footnote{Having in mind
this trick one can say that the matrix element is gauge-invariant in the
$k_T$-factorization approach. However the polarization vectors of
incoming gluons $q_1$ and $q_2$ are not arbitrary taken but fixed as
$-q_{1T}^\mu/y$ and $-q_{2T}^\nu/x$, respectively (see \cite{GLR} for
more detail).}.  Both these equations for $d_{\mu\nu}$  can be used
but for the form (\ref{trans1}) one has to modify the gluon vertex
slightly (to account for several ways of gluon emission -- see ref.
\cite{3}):
\begin{equation}
\label{geff}
\Gamma_{eff}^{\nu}\ =\
\frac{2}{xys}\ \left[(xys - q_{1T}^2)\,q_{1T}^{\nu} - q_{1T}^2
q_{2T}^{\nu} + 2x\,(q_{1T}q_{2T})\,p_B^{\nu}\right]\ .
\end{equation}
As a result the colliding gluons can be treated as aligned ones and
their polarization vectors are directed along the transverse momenta.
Ultimately, the nontrivial azimuthal correlations must arise between
the transverse momenta $p_{1T}$ and $p_{2T}$ of the heavy quarks.

From the formal point of view there is a danger to loose the gauge
invariance in dealing with the off mass shell gluons. Say, in the
covariant Feynman gauge the new graphs (similar to the "bremsstruhlung"
from the initial or final quark line, as it is shown in Fig.~6d) may
contribute in the central plato rapidity region. However this is not the
fact. Within the "semihard" accuracy, when the function $\varphi(x,q^2)$
collects the terms of the form $\alpha_s^k(\ln q^2)^n(\ln (1/x))^m$ with
$n+m\ge k$, the triple gluon vertex (\ref{geff}) includes effectively all the
leading logarithmic contributions of the Fig.~6d type \cite{BFKL,8}.
For instance, the upper part of the graph shown in Fig.~6d corresponds
in terms of the BFKL equation to the $t$-channel gluon reggeization.
Thus the final expression is gauge invariant (except a small,
non-logarithmic, $O(\alpha_s)$ corrections) \footnote{Taking the gluon
polarization vector as $-q_{1T}^\mu/y$ we can completely eliminate the
leading logarithm terms arising from Fig.~6d.}.

Although the situation considered here seems to be quite opposite to the
parton model there is a certain limit \cite{3}, in which our formulae
can be transformed into parton model ones, namely when the quark
transverse momenta, $p_{1,2T}$, are much larger than the gluon ones,
$q_{1,2T}$.

\section{Qualitative difference between $k_T$-factorization approach
and parton model}

Eq.~(\ref{spp}) enables to calculate straightforwardly all distributions
concerning one-particle or pair production. To illustrate the difference
between our approach and the conventional parton model we present first
of all the results of calculations of charm production ($m_c$ = 1.4~GeV
\cite{Nar,BBB}) with high $p_T$ at three energies, $\sqrt{s}$ = 1~TeV,
10~TeV and 10$^3$~TeV and with the same value of
\begin{equation}
x_T\ =\ \frac{2 p_T}{\sqrt{s}}\ =\ 0.02\ ,
\end{equation}
i.e. $p_T$ = 10~GeV, 100~GeV and 10$^4$~Gev for the above energies.

We will use QCD scales $\mu_F = m_T = \sqrt{m^2_c + p^2_T}$ and
$\mu_R = m_c$, i.e. fixed coupling with $N_f$ = 3 and $\Lambda$ = 248~
MeV \cite{GRV}.

However there exists a problem coming from the infrared region, because
the functions $\varphi (x,q^2_2)$ and $\varphi (y,q^2_1)$ are unknown
at the small values of $q^2_{1,2}$. Moreover, for the realistic gluon
structure functions the value $\varphi(x,q^2)$ is positive in the
small-$x$ region and negative in the large-$x$ region. The boundary
between two regions, where $\varphi (x,q^2)$ =0, depends on the $q^2$,
therefore the relative contributions of these regions is determined by
the characteristic scale of the reaction, say, by the transverse
momentum $p_T$.

To avoid this additional problem, we present the numerical
calculations both in the $k_T$-factorization approach and in the
LO parton model, using the "toy" gluon distribution
\begin{equation}
\label{toy}
xG(x,q^2)\ =\ (1-x)^5 \ln(q^2/Q_0^2)\ ,
\end{equation}
for $q^2 > Q_0^2$, and $xG(x,q^2) = 0$ for $q^2 < Q_0^2$, with
$Q_0^2$ = 1~GeV$^2$.

After the calculations of charm production cross sections using, say,
the VEGAS code \cite{Lep} (that is the optimized Monte Carlo program
for integrating the multidimentional expressions) and with Eqs.~(12),
(22), (32) one can see, that the values $d \sigma /d x_T$ at $x_T$ =
0.02 are about 4--5 times larger in the $k_T$-factorization approach
than in the LO parton model, Eq.~(1). Really this difference should
not be considered as very large because, as it was discussed, an
essential part of NLO and NNLO corrections is already included in the
$k_T$-factorization, and it is known that the sum of the LO and NLO
contributions in the parton model is about 2-3 times larger than the
LO contribution \cite{Liu}.

To show the most important values of the variables $q_{1,2T}$
in Eq.~(\ref{spp}), as well as the kinematical region producing the main
difference with the conventional parton model, we plot by dots in
Fig.~7 the results of the $k_T$-factorization approach with the
restrictions $\vert q_{1,2T} \vert \leq q_{max}$, as a function of
$q_{max}$. The running coupling is fixed as $\alpha_S(m^2_c)$ instead
of $\alpha_S(q^2_{1,2})$ in Eq.~(\ref{spp}). The dashed lines in
Fig.~7 are the conventional parton model predictions with QCD scales
$\mu_F = \sqrt{m_c^2 + p_T^2}$, and $\mu_R = m_c$. One can see that
the $k_T$-factorization predictions exceed the parton model results
when $q_{max} \geq p_T$.

\begin{figure}
\begin{center}
\mbox{\epsfig{file=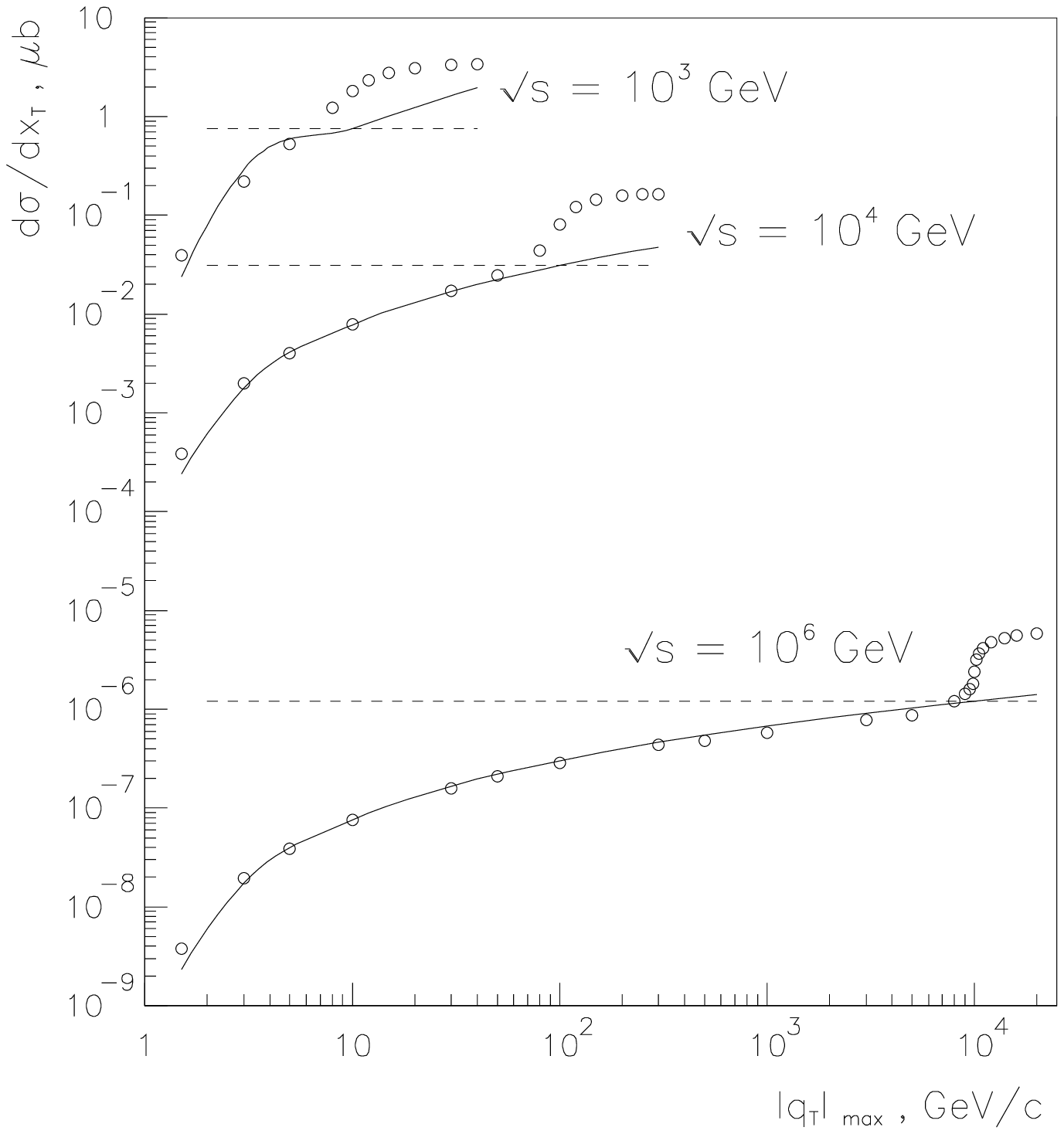,width=0.60\textwidth}} \\
Fig. 7. The values of $d\sigma / dx_T$ for charm
hadroproduction at $x_T$ = 0.02 in $k_T$-factorization approach as a
function of upper limits of integrals in Eq.~(22) (points); the
conventional parton model values (dashed lines) and the values of
right-hand side of Eq.~(34) (solid curves).
\end{center}
\end{figure}

Let us check now that the $k_T$-factorization predictions really
coincide with the parton model ones when the quark momenta,
$p_{1,2T}$, are much larger than the gluon momenta, $q_{1,2T}$ \cite{3}.
However we have to compare them at the same values of structure functions.
When we submit Eq.~(\ref{spp}) to the conditions
$\vert q_{1,2T} \vert \leq q_{max}$ and neglect the $q_{1,2T}$ dependence
in the matrix element, we recover the parton model approximation, and get
the result proportional to $xG(x,q^2_{max}) \cdot yG(y,q^2_{max})$
due to the direct consequence of Eq.~(\ref{xG}) \cite{Kwi}
\begin{equation}
xG(x,q^2)\ =\ xG(x,Q_0^2) + \frac{1}{4\sqrt{2} \pi^3}
\int_{Q_0^2}^{q^2} \varphi(x,q_1^2) dq_1^2\ .
\end{equation}

At the same time the original parton model yields the result
proportional to $xG(x,\mu_F^2) \cdot yG(y,\mu_F^2)$ with
$\mu_F = \sqrt{m_c^2 + p_T^2}$. Hence we expect the parton model
to coincide with our calculations for the gluon distribution
Eq.~(\ref{toy}), $p_T \gg m_c$ and $\vert q_{1,2T} \vert \leq q_{max}$
after multiplying by an appropriate factor:
\begin{equation}
\label{f}
\frac{d \sigma}{d x_T} \left \vert _{q_{iT} \ll p_{iT}}\ =\
\frac{d \sigma}{d x_T} \right \vert _{PM}
\left (\frac {\ln(q^2_{max}/Q_0^2)}{\ln(p_T^2/Q_0^2)} \right )^2  \;.
\end{equation}

\begin{figure}
\begin{center}
\mbox{\epsfig{file=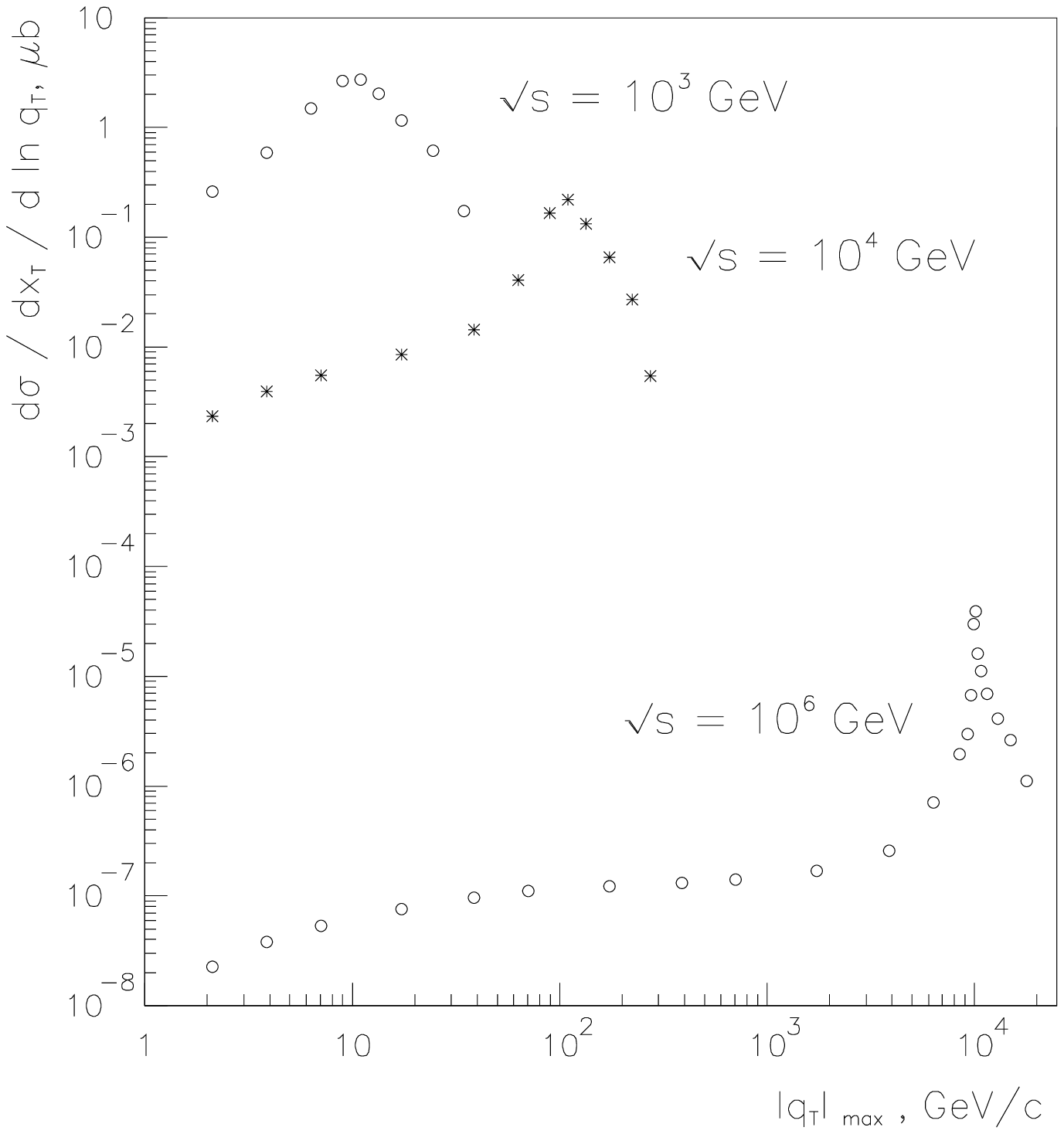,width=0.60\textwidth}} \\
Fig. 8. The values of $d\sigma / dx_T/d \ln q_{max}$ for charm
hadroproduction at $x_T$ = 0.02 in $k_T$-factorization
approach as a function of upper limits of integrals in
Eq.~(22) (points).
\end{center}
\end{figure}

The right-hand side of Eq.~(\ref{f}) presented in Fig.~7 by solid
curves shows a good agreement with $k_T$-factorization approach (open
dots) even when $q_{max}$ is only slightly smaller (at the highest
energy) than $p_T$. The same values $d \sigma / d x_T$, as in Fig.~7,
but differentiated with respect to $\ln{q_{max}}$ are presented in
Fig.~8 for $k_T$-factorization approach. It exhibits, especially at
the high energies, the logarithmic growth with $q_{max}$, until the
value $q_{max} \sim p_T$. There is the narrow peak in this region,
which provides about 70-80 \% of the $d \sigma / d x_T$ cross section
integrated over the whole $q_{1,2T}$ phase space. Its physical nature
seems to be quite clear. There are two kinematical regions for
$t$-channel and $u$-channel diagrams, Fig.~6~a,b, giving
comparatively large contributions to $d \sigma / d x_T$ for the high
energies and high $p_T$ heavy quark production. One of them comes from the
conventional parton model kinematics when both initial gluon transverse
momenta, $q_{1,2T}$, are very small compared to $p_{1,2T}$, see
Fig.~9a.  Another large contribution appears from the region where,
say, $q_{1T} \sim p_{1T}$, whereas $q_{2T}$ and $p_{2T}$ are
comparatively small, as it is shown in Fig.~9b. In this case the
quark propagator,
$1/(\hat{p_1} - \hat{q_1} - m_Q) = 1/(\hat{q_2} - \hat{p_2} - m_Q)$,
is close to the mass shell and gives rise to the narrow peak shown in
Fig.~8. The general smallness of such processes comes from
high-virtuality gluon propagator in Fig.~9b, and it is of the
same order as in the case of Fig.~9a, despite the diagram Fig.~9a
is suppressed by the fermion propagator for the large rapidity difference
between the two produced heavy quarks.
\begin{figure}
\begin{center}
\mbox{\epsfig{file=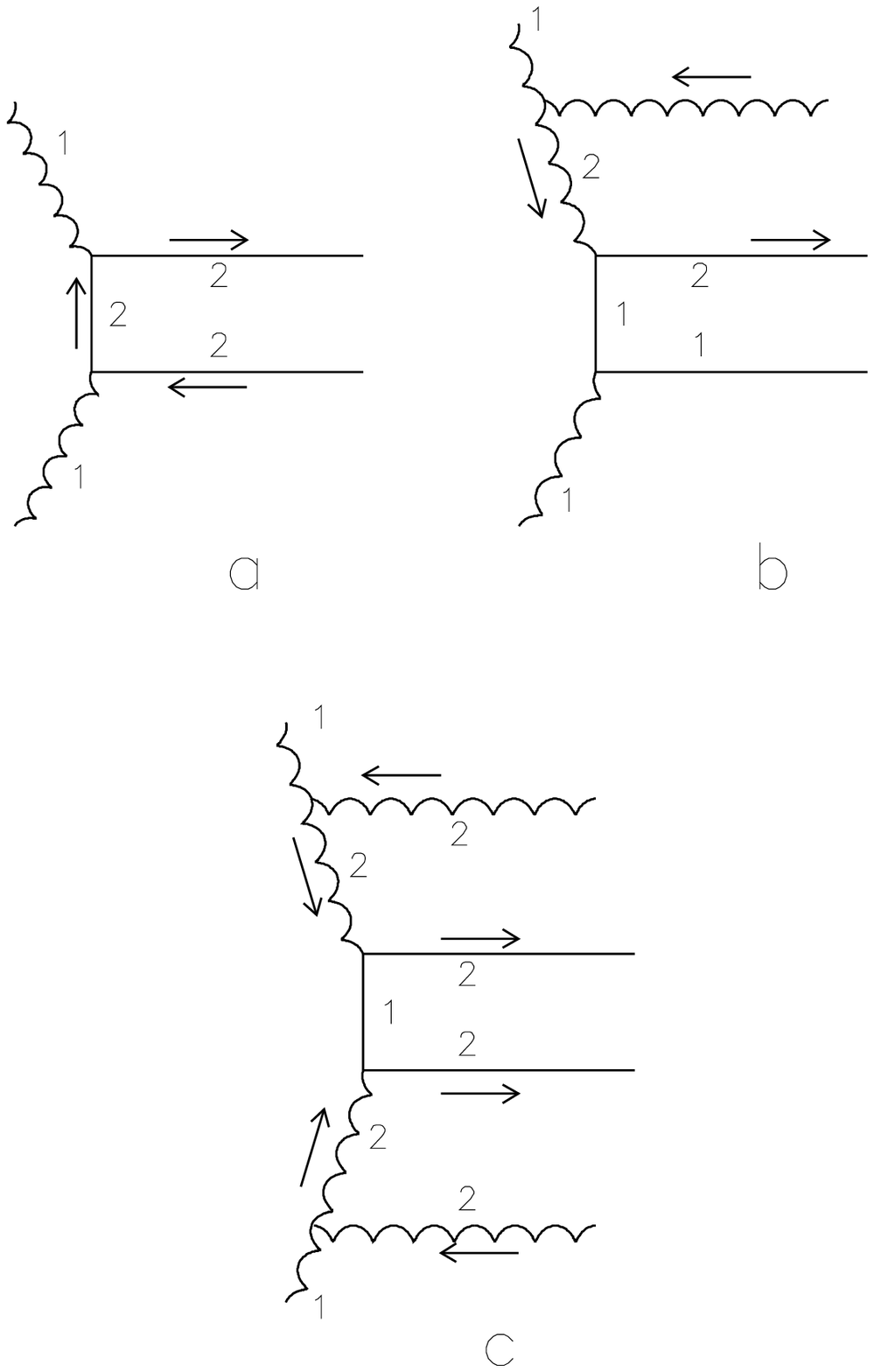,width=0.60\textwidth}} \\
Fig. 9. The diagrams which are important in the case of one-particle
distributions of heavy quark with large $p_T$. High $p_T$ flows are
shown by arrows. The situation similar to the LO parton model (a).
The case, possible in NLO parton model, large $p_T$ of the quark is
compensated by hard gluon and the fermion propagator is near to mass
shell (b).  The numerically small contribution, when large transverse
momenta of heavy quarks are compensated by two hard gluons, whereas
the fermion propagator is near to mass shell (c).
\end{center}
\end{figure}

The diagram shown in Fig.~9b can be considered \cite{1} as one of the
most important contribution to the NLO parton model in the high
energy limit, because of spin-one gluon exchange in the $t$-channel. Due
to these factors, combinatorics and interference between
diagrams, the resulting $k_T$-factorization contribution
to the peak at $q_T \sim p_T$ in the total
$d \sigma/ d x_T$ value in Fig.~8 is several times larger than the LO parton
model contribution. In the Table~1 we present the calculated ratios of the
total heavy quark pair production cross section, $R_{tot}$, and the
ratio of $d \sigma /d x_T$, $R(x_T)$ at $x_T$ = 0.02 (integrated over
rapidity).

\begin{table}
\caption{The ratios of $c\bar c$ pair production in
$k_T$-factorization approach and in LO parton model}

\begin{center}
\vskip 20 pt
\begin{tabular}{||c|r|r|r|r|r||} \hline\hline

$\sqrt s$, TeV & 0.3 & 1   & 10  & 100 & 1000 \\ \hline

$R_{tot}$       & 4.0 & 4.0 & 4.0 & 3.9 & 3.9 \\

$R(x_T = 0.02)$  & 3.4 & 4.5 & 5.5 & 5.4 & 5.2 \\
\hline\hline
\end{tabular}
\end{center} \end{table}
\vskip 10 pt

One can see that at fixed $x_T$ the relative contribution of the
discussed peaks at first increases due to increase of the phase space.
This contribution is saturated at the high enough energy
($\sqrt{s} \sim 10$ TeV).
After this the relative contribution of the LO parton
model increases logarithmically with $p_T$, and have to dominate at
the extremely high energies and $p_T$, in academical asymptotic.

These results are confirmed by those presented in Fig.~10, where we
reproduce the data from Fig.~7 for LO parton model and
$k_T$-factorization approach with the condition
$\vert q_{1,2T} \vert \leq q_{max}$. For comparison we show by stars
the $k_T$-factorization predictions for all values $q_{1T}$ with the
only condition $\vert q_{2T} \vert \leq q_{max}$, versus $q_{max}$. The
last results are above the LO parton model even at not too large
$q_{2T}$ because the peaks, discussed above, are included into the integral
over $q_{1T}$.

\begin{figure}
\begin{center}
\mbox{\epsfig{file=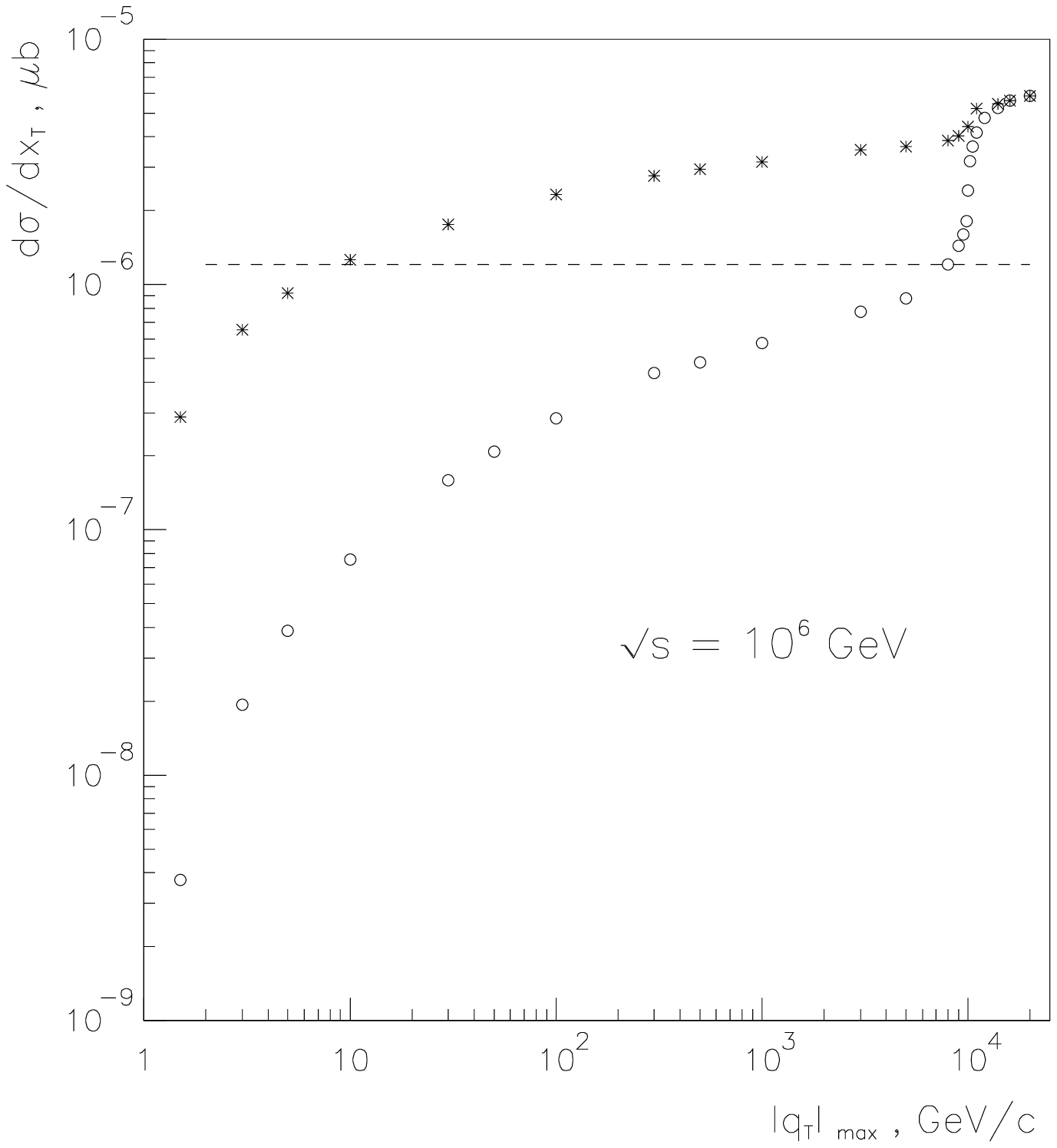,width=0.60\textwidth}} \\
Fig. 10. The same as in Fig.~7 (points and dashed curves) together
with the values of $d \sigma /d x_T$ (stars) for the case when only
$q_{2T}$ upper limit integration is bounded by $q_{max}$.
\end{center}
\end{figure}

Let us note that the calculation of $d \sigma /d x_T$ at $\sqrt{s}$ = 10 TeV,
$x_T$ = 0.02, and with restriction $\vert q_{1,2T} \vert \geq p_T/2$
(see Fig.~9c) gives only about 2\% of the total value of $d \sigma /d
x_T$.

The essential values of $q_{1,2T}$ increase in our calculations when
the transverse momentum, $p_T$ of the detected $c$-quark increases.
At the high initial energy $q_{1,2T} \sim p_T$. In the $k_T$ kick
language it means that the $\langle k^2_T\rangle$ value also
increases.

\section{Total cross sections and one-particle distributions}

Eq.~(22) enables us to calculate straightforwardly all distributions
concerning one-particle or pair production. One-particle
calculations as well as correlations between two produced heavy quarks
can be easily done using, say, the VEGAS code \cite{Lep}.

However there exists a principle problem coming from the infrared
region.  Since the functions $\varphi (x,q^2_2)$ and $\varphi
(y,q^2_1)$ are unknown at small values of $q^2_2$ and $q^2_1$, i.e.
in nonperturbative domain we use Eq.~(33) and rewrite the integrals in
the Eq.~(22) as
%\newpage
$$\int d^2 q_{1T} d^2 q_{2T} \delta (q_{1T} + q_{2T} -
p_{1T} - p_{2T}) \frac{\alpha_s(q^2_1)}{q_1^2} \frac{\alpha_s
(q^2_2)}{q^2_2} f_g(y,q_{1T},\mu) f_g (x,q_{2T},\mu) \vert
M_{QQ}\vert^2 = $$
\begin{equation}
\label{int}
= (4\sqrt{2}\,\pi^3 \alpha_s (m^2_T) )^2 \, xG(x,Q^2_0)\, yG(y,Q^2_0)\,
T^2(Q_0^2,\mu^2)\, \left (\frac{\vert M_{QQ} \vert^2}{q^2_1 q^2_2}
\right)_{q_{1,2}\rightarrow 0} \; +
\end{equation}
$$ + \; 4\sqrt{2}\,\pi^3 \alpha_s (m^2_T) xG(x,Q^2_0)\,T(Q_0^2,\mu^2)\,
\int^{\infty}_{Q^2_0} \, d q^2_{1T}\, \delta (q_{1T} - p_{1T} -
p_{2T})\,\, \times $$
$$ \times \, \frac{\alpha_s (q^2_1)}{q^2_1} f_g(y,q_{1T},\mu)
\left (\frac{\vert M_{QQ} \vert^2}{q^2_2}
\right)_{q_2\rightarrow 0} \; + $$
$$ + \; 4\sqrt{2}\,\pi^3 \alpha_s (m^2_T)\, yG(y,Q^2_0)\,T(Q_0^2,\mu^2)\,
\int^{\infty}_{Q^2_0} \, d q^2_{2T}\, \delta
(q_{2T} - p_{1T} - p_{2T})\, \, \times $$
$$ \times \, \frac{\alpha_s (q^2_2)}{q^2_2} f_a(x,q_{2T},\mu)
\left (\frac{\vert M_{QQ}
\vert^2}{q^2_1} \right)_{q_1 \rightarrow 0} \; + $$
$$ + \; \int^{\infty}_{Q^2_0} \, d^2 q_{1T}
\int^{\infty}_{Q^2_0} \, d^2 q_{2T}\,
\delta (q_{1T} + q_{2T} - p_{1T} - p_{2T}) \, \times $$
$$ \times \,\frac{\alpha_s(q^2_1)}{q_1^2} \frac{\alpha_s (q^2_2)}{q^2_2}
f_g(y,q_{1T},\mu) f_g (x,q_{2T},\mu) \vert M_{QQ}\vert^2 \;, $$
where the unintegrated gluon distributions $f_g(x,q_T,\mu)$ are taken
from Eqs.~(14) and (17). In the numerical calculations we use the
values $\mu^2 = \hat{s}$ and $\mu^2 = \hat{s}/4$.

The first contribution in Eq.~(35) with the matrix
element averaged over directions of the two-dimensional vectors $q_{1T}$
and $q_{2T}$ is exactly the same as the conventional LO parton model
expression. It is multiplied by the 'survival' probability
$T^2(Q_0^2,\mu^2)$ not to have transverse momenta $q_{1T}, q_{2t} >
Q_0$.  The QCD scales are $\mu_R^2 = m^2_T$ and $ \mu_F^2 = Q_0^2$
(the same value as in Eq.~(33), we assume $Q_0^2$ = 1~GeV$^2$)
The sum of the produced
heavy quark momenta is taken to be  exactly zero here.

The next three terms contain the corrections to the parton model due
to gluon polarizations, virtualities and transverse momenta in the
matrix element. The relative contribution of the corrections strongly
depends on the initial energy. If it is not high enough, the first term
in Eq.~(35) dominates, and all results are similar to the
conventional LO parton model predictions. In the case of very high
energy the opposite situation takes place, the first term in Eq.~(35)
can be considered as a small correction and our results differ from
the conventional ones. It is necessary to note that the absolute as
well as the relative magnitudes of all the pieces in Eq.~(35) strongly depend
on the T-factor (i.e., when we use Eq.~(17) or Eq.~(12)).

Before the numerical comparison it is necessary to remind that the NLO
parton model actually results only in a normalization factor in the
case of one-particle distributions, the shapes of LO and LO+NLO
distributions are almost the same \cite{NDE,Beer,Beer1,MNR1}, see
Fig.~2. This means that we can calculate the K-factor Eq.~(7), say,
from the results for the total production cross sections, and
restrict ourselves only to the LO calculations of $p_T$ or rapidity
distributions multiplying them by the K-factors.

The numerical values of the K-factors depend \cite{Liu} on the
structure functions used, quark masses, QCD scales and the initial energy,
the renormalization scale $\mu_R$ being especially
important. This seems to be evident, because the LO contribution is
proportional to $\alpha_S^2$, whereas the NLO contribution is
proportional to $\alpha_S^3$. However the structure of Eq.~(2) is
more important at the high energies, when small $\rho$ values dominate.
At $\rho \to 0$ the functions $f^{(1)}_{gg}$ and
$\bar{f}^{(1)}_{gg}$ have the constant limits \cite{1},
$f^{(1)}_{gg}(\rho \to 0) \approx 0.1$ and
$\bar{f}^{(1)}_{gg}(\rho \to 0) \approx -0.04$, while
$f^{(0)}_{gg}(\rho \to 0) \to 0$. Therefore due to Eq.~(2) the K factor
value at the high energies is mainly determined by the ratio $\mu / m_Q$.

First of all let us present the role of the $T$-factors, Eq.~(14). In
Fig.~11 we show the ratios of the last term of Eq.~(35) with the
$T$-factors in both gluon distributions $f_g(x,q_{iT},\mu)$ to the
same values calculated for $T_g(x,q_{1T},\mu)=1$ in one unintegrated
gluon distribution. The ratios are plotted as functions of $q_{1T}$
for the cases of charm and beauty production at $\sqrt s=14$~TeV and
$\mu^2=\hat{s}$. The heavy quark transverse momenta are fixed to be
20~GeV/c. In both cases the factors $T_a(q_{1t})$ are rather small at
small $q_{1T}$ and $T_g(x,q_{1T},\mu)\to 1$ at $p_T \gg q_{1T}$.

\begin{figure}
\begin{center}
\mbox{\epsfig{file=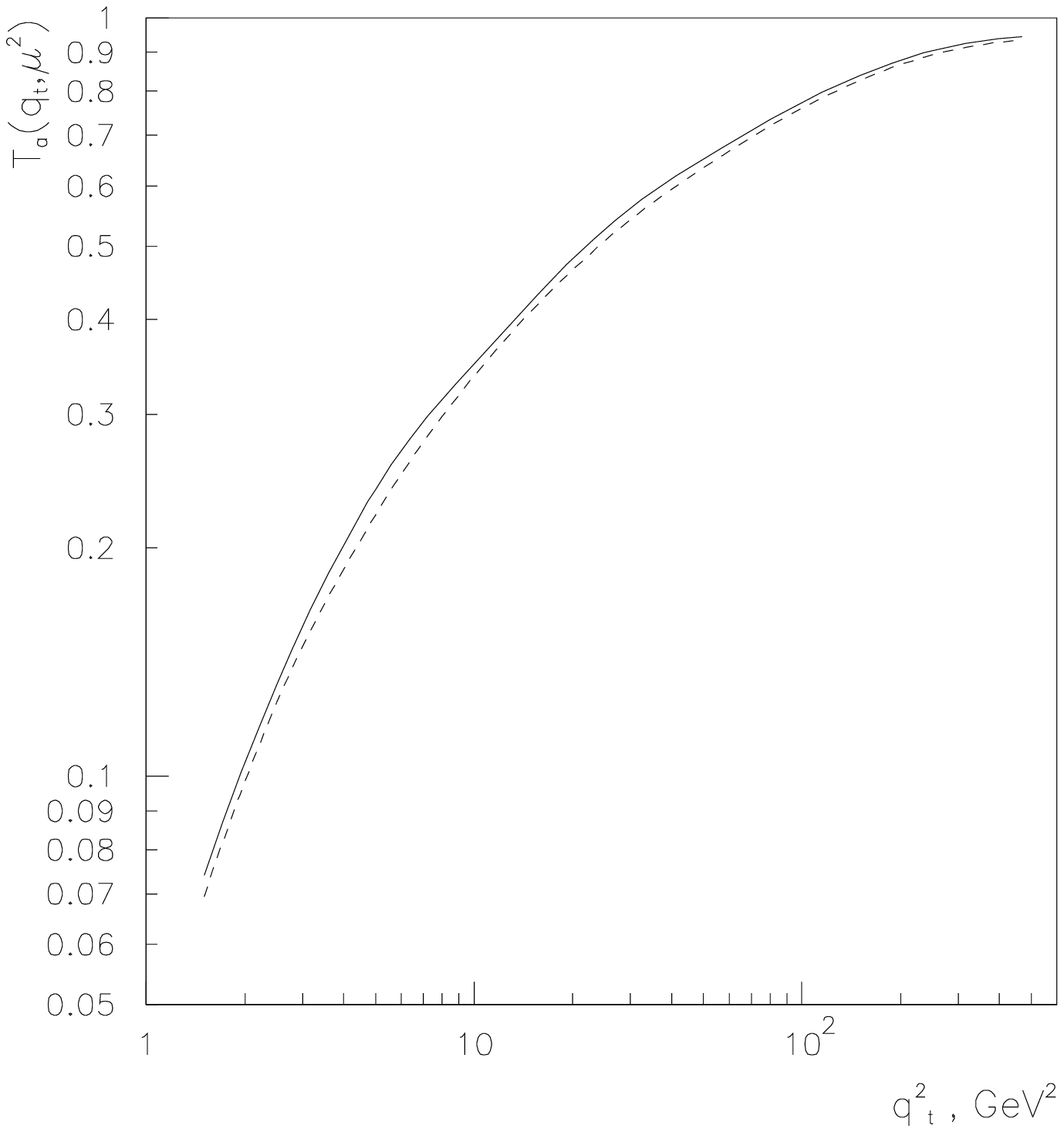,width=0.60\textwidth}} \\
Fig. 11. The role of $T$-factor, Eq. (14) in the calculation of charm
(solid curve) and beauty (dashed curve) production with $p_T=20$~GeV/c
at $\sqrt{s}$ = 14 TeV and $\mu^2 = \hat{s}$.
\end{center}
\end{figure}

Now let us compare the numerical results predicted by the parton model
and by the $k_T$-factorization approach.

\begin{figure}
\begin{center}
\mbox{\epsfig{file=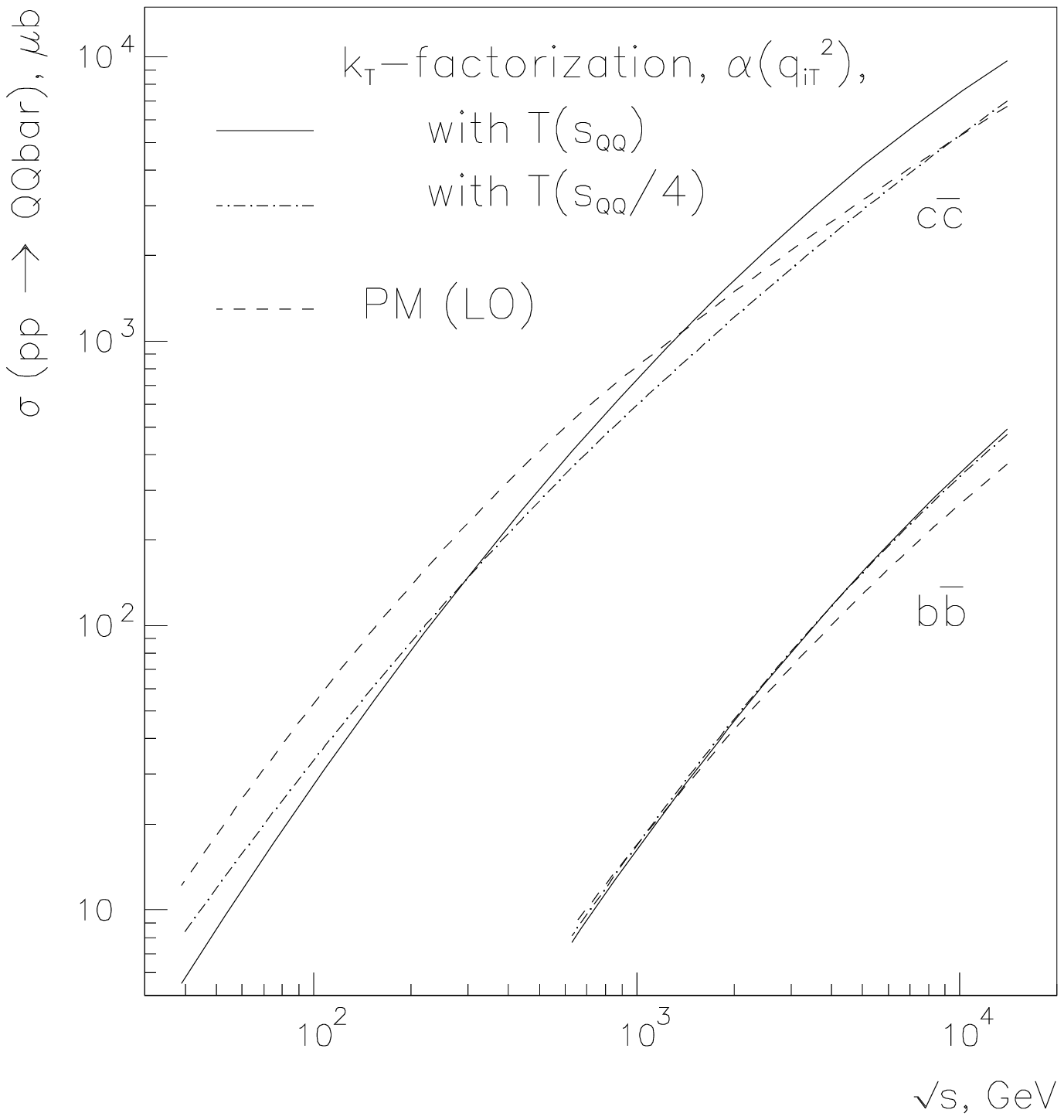,width=0.60\textwidth}} \\
Fig. 12. The total cross sections for charm and beauty hadroproduction
in the $k_T$-factorization approach with unintegrated gluon distribution
$f_g(x,q_T,\mu)$ given by Eq.~(17), for $\mu^2$ values in Eq.~(14)
equal to $\hat{s}$ (solid curves), and $\hat{s}/4$ (dash-dotted
curves), and in the LO parton model (dashed curves).
\end{center}
\end{figure}

The energy dependences of the total cross sections of $c\bar{c}$ and
$b\bar{b}$ pair production are presented in Fig.~12. As was
mentioned, at comparatively small energies the first term in Eq.~(35)
dominates and the results of the $k_T$-factorization approach should
be close to the LO parton model prediction. Actually the first
results are even smaller due to the presence of the $T$-factor in
Eq.~(17).  However the $k_T$-factorization approach yeilds a
stronger energy dependence than the LO parton model both for
$c\bar{c}$ and $b\bar{b}$ production. This can be explained by
additional contributions appearing at very high energies in the
$k_T$-factorization approach when the transverse momentum of a heavy
quark is compensated by the nearest gluon, see \cite{our}

\begin{figure}
\begin{center}
\centerline{\epsfig{figure=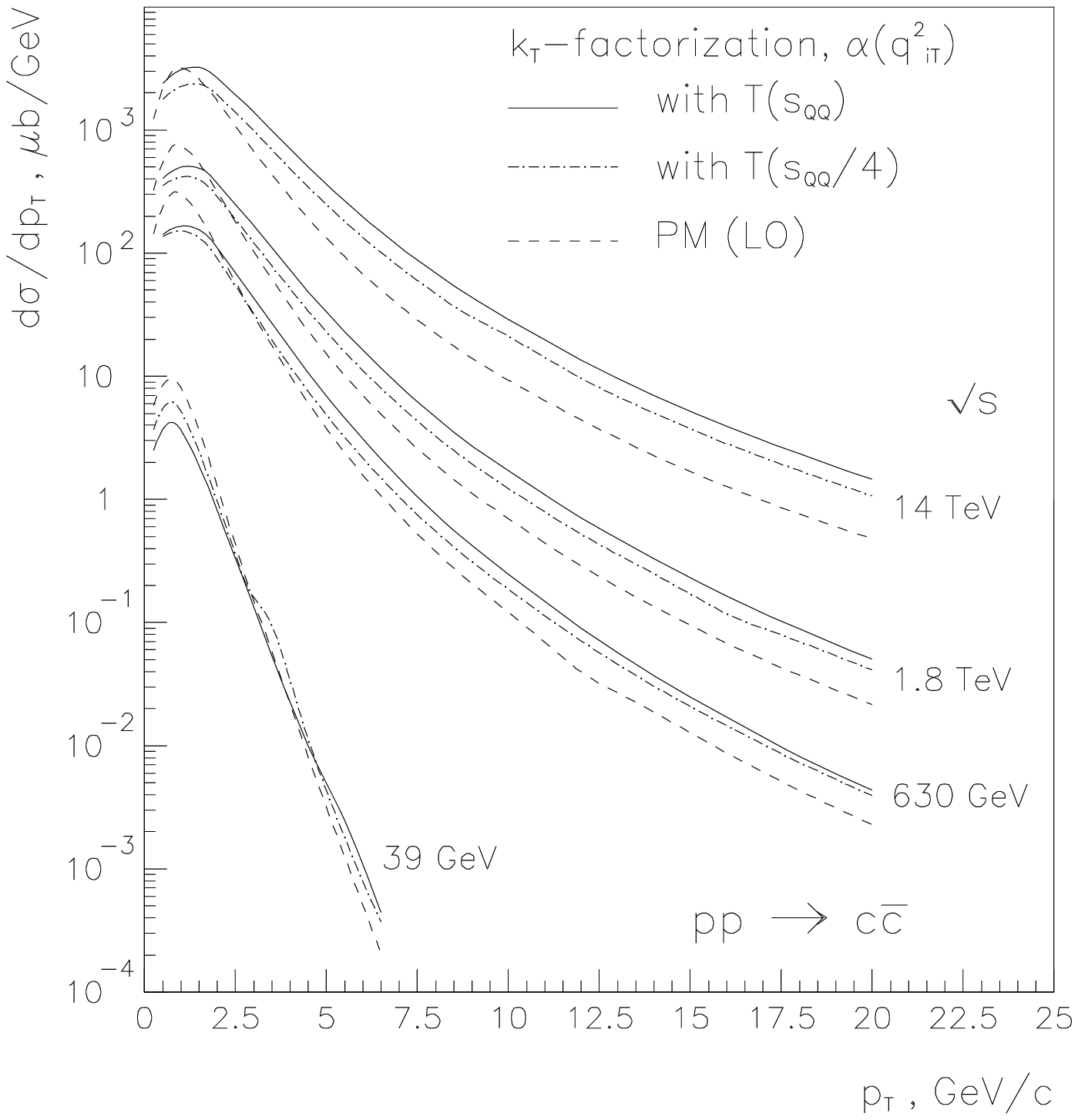,width=0.50\textwidth,clip=}
            \hspace{0.3cm}
            \epsfig{figure=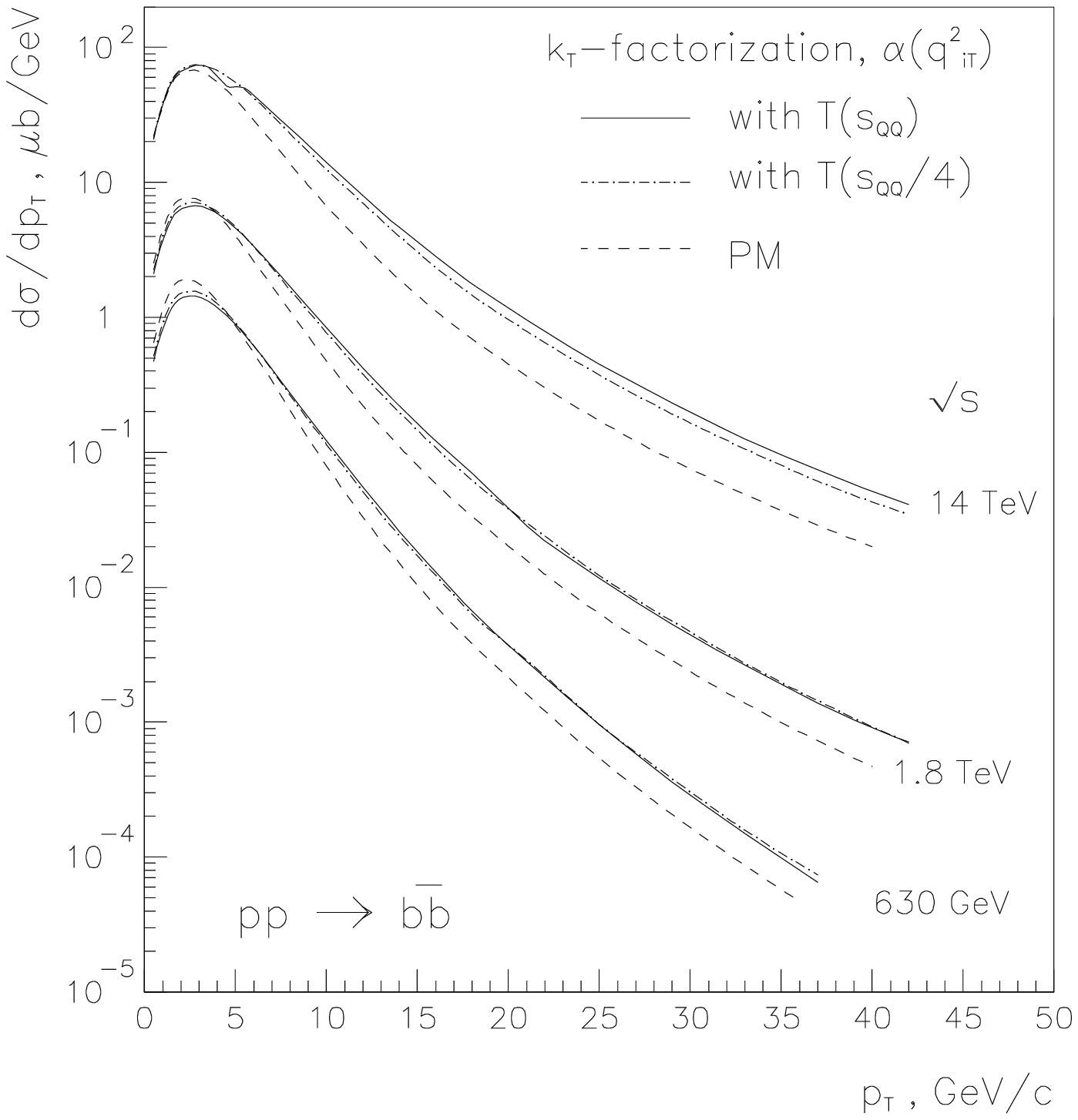,width=0.50\textwidth,clip=}}
Fig. 13. $p_T$-distributions of $c$-quarks (a) and $b$-quarks (b)
produced at different energies. Dashed curves are the results of the LO
parton model. Solid curves are calculated with the unintegrated gluon
distribution $f_g(x,q_T,\mu)$ given by Eq.~(17), for
$\mu^2$ values in Eq.~(14) equal to $\hat{s}$ and dash-dotted curves
are calculated for $\mu^2 = \hat{s}/4$.
\end{center}
\end{figure}

One-particle $p_T$ distributions, $d \sigma /dp_T$, calculated
in the $k_T$-factorization approach and in the LO
parton model are presented in Fig.~13. In all cases the
$k_T$-factorization predicts broader distributions. The average
values of $p_T$ of the produced heavy quarks are rather different in
these two approaches, as one can see from Table 2.

%\newpage
\vskip 10 pt
\begin{center}
{\bf Table 2}
The average values of charm and beauty quark transverse momenta
$<\!p_T\!>$ (in GeV/c) in the $k_T$-factorization approach with
$\mu^2 = \hat{s}$ and in the LO parton model.
\vskip 20 pt
\begin{tabular}{|c|r|r|r|r|} \hline

 & \multicolumn{2}{c|}{LO parton model} &
\multicolumn{2}{c|}{$k_T$-factorization}  \\ \hline

$\sqrt{s}$ & $c\bar{c}$ & $b\bar{b}$ & $c\bar{c}$ & $b\bar{b}$
\\ \hline

14 TeV   & 1.78 & 4.53 & 2.23 & 5.47 \\ \hline

1.8 TeV  & 1.48 & 3.96 & 1.91 & 4.54  \\  \hline
\end{tabular}
\end{center}
\vskip 10 pt

This seems to be very natural, because, contrary to the LO
parton model, a large $p_T$ of the one heavy quark can be compensated not
only by the $p_T$ of the other heavy quark but also by neighbour gluons
emission, Fig.~9b.

The rapidity distributions of produced heavy quarks presented in
Fig.~14 show that the main part of the difference between the
$k_T$-factorization approach and the LO parton model comes from the
central region.

\begin{figure}
\begin{center}
\centerline{\epsfig{figure=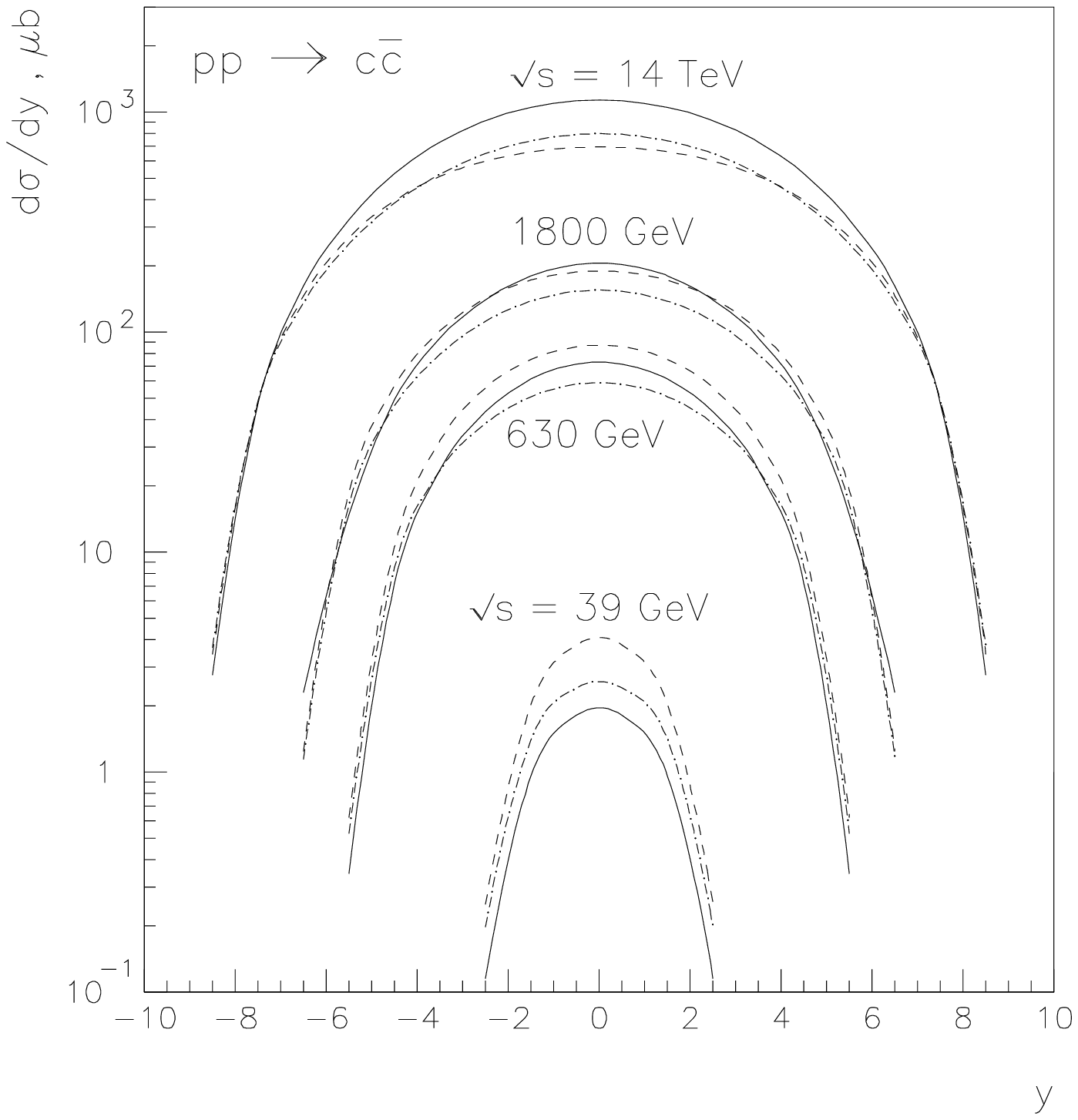,width=0.50\textwidth,clip=}
            \hspace{0.3cm}
            \epsfig{figure=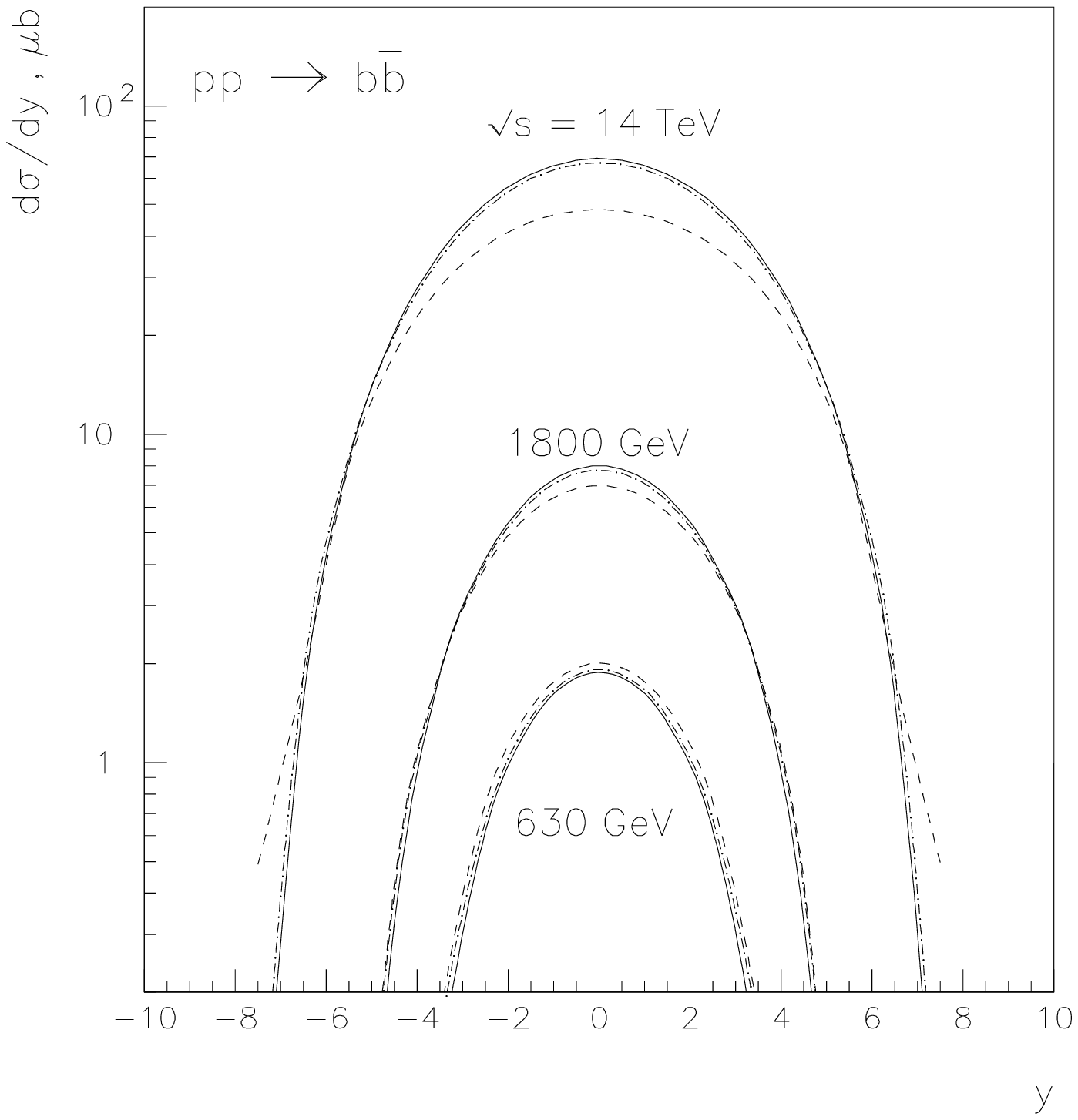,width=0.50\textwidth,clip=}}
Fig. 14. Rapidity distributions of $c$-quarks (a) and $b$-quarks (b)
produced at different energies. Dashed curves are the results of the LO
parton model. Solid curves are calculated with the unintegrated gluon
distribution $f_g(x,q_T,\mu)$ given by Eqs.~(17), for $\mu^2$ values
in Eq.~(14) equal to $\hat{s}$ and dash-dotted curves are calculated
for $\mu^2 = \hat{s}/4$.
\end{center}
\end{figure}

\section{Two-particle correlations}

We saw from the previous section that there is not so large difference
in our results for the total cross sections and one-particle
distributions obtained in the $k_T$-factorization and in the LO parton
model. The predictions of the NLO parton model for these quantities
differ from the LO parton model only by a normalization factor of
2-2.5 \cite{NDE,Beer,Beer1,MNR1}. Hence the difference between our
predictions and the NLO parton model should be small.

The calculations of two-particle correlations in different approaches
are more informative. The simplest and/or most informative correlation
here is the distribution of the total transverse momentum of the
produced heavy quark pair, $p_{pair}$. In the LO parton model
$p_{pair} = p_{1T} + p_{2T} = q_{1T} + q_{2T}$, therefore if $q_{1T}
= q_{2T} = 0$, then $d \sigma /d p_{pair}$ is a $\delta$-function
with $p_{pair} = 0$.  Thus the $p_{pair}$ distributions provide the
direct information about the transverse momentum distribution of the
incident partons.

It is clear that if $q_{iT} \ll p_{iT}$, then the distributions in
$p_{pair}$ should be narrower in comparison with the one-particle
$p_T$ distributions. In this case the Weizsaecker-Williams approximation
should be valid and one can believe that the parton model reflects the
real dynamics of the interaction. In the opposite case,
$q_{iT} \sim p_{iT}$, the large transverse momentum of the produced
heavy quark can be compensated not by the other quark, but by a
high-$p_T$ gluon. We have shown \cite{our} (see Sect.~4), that
about 70-80\% of the total cross section of high-$p_T$ quark production
at high energies originates from such processes, when the heavy quark
propagator is close to the mass shell.

We calculate the values of $d \sigma /d p_{pair}$ for charm (a) and
beauty (b) production in the $k_T$-factorization approach using the
unintegrated gluon distribution Eqs.~(14), (17) with scale values
$\mu^2 = \hat{s}$ and $\mu^2 = \hat{s}/4$ (the last value only for
$\sqrt s =14$~Tev).  Our results for pair production at different
initial energies are shown by solid curves in Fig.~15. For the
comparison we present by dashed curves the one-particle
$p_T$-distributions taken from Fig.~13, obtained in the same
$k_T$-factorization approach and with the same $T$-factor.  As we put
$Q_0^2$ = 1 GeV$^2$ in Eq.~(35), we can not distinguish between the
initial gluons with $q_T$ equal to, say, 0.1 GeV/c and 0.9 GeV/c, so
our first bin in the $d \sigma /d p_{pair}$ distribution has the
width 2~GeV/c which explains some irregular behaviour of the solid
curves at the small $p_T$.  Naturally, all the solid and dashed
curves are equally normalized at the same energy.

\begin{figure}
\begin{center}
\centerline{\epsfig{figure=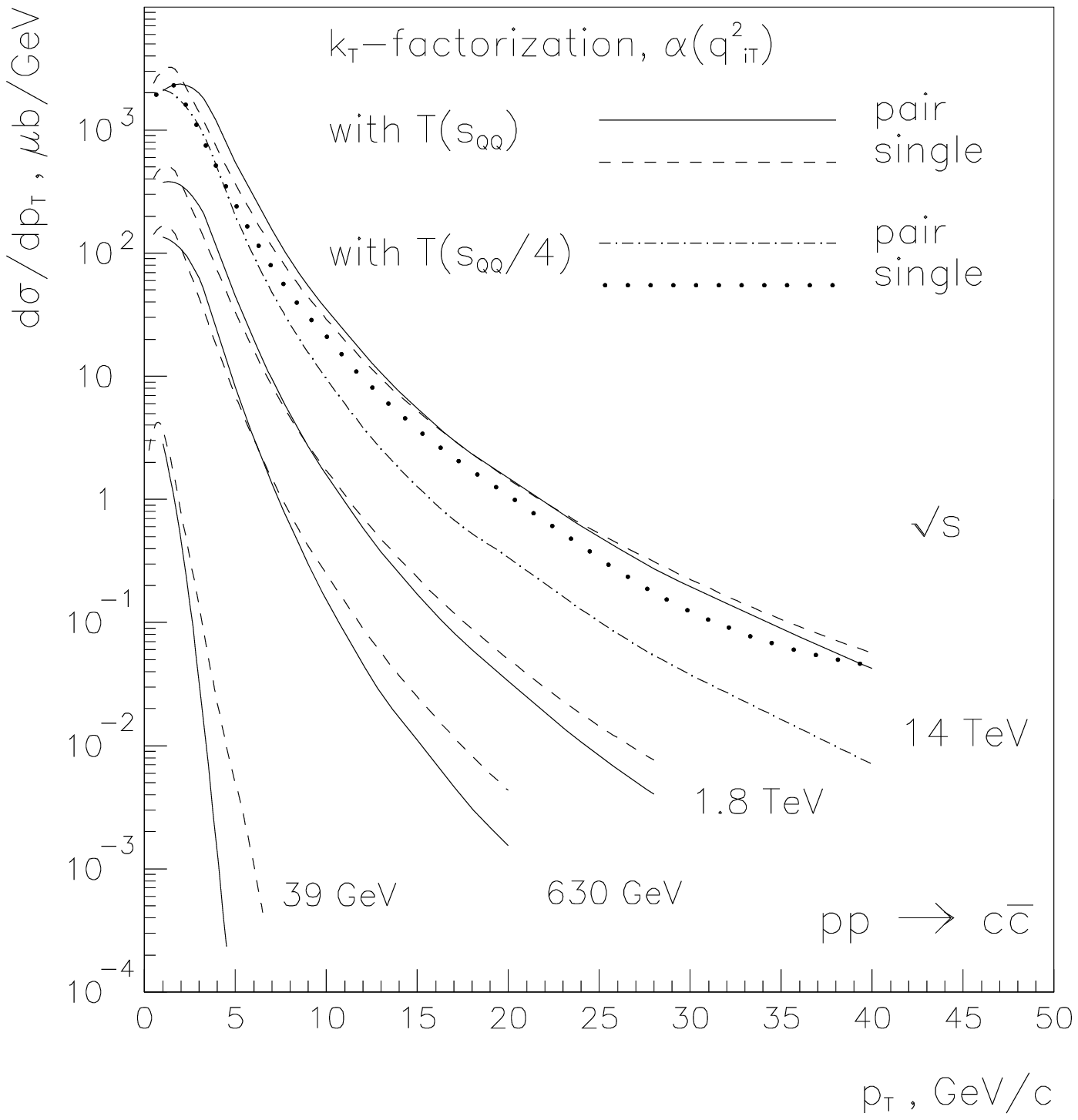,width=0.50\textwidth,clip=}
            \hspace{0.3cm}
            \epsfig{figure=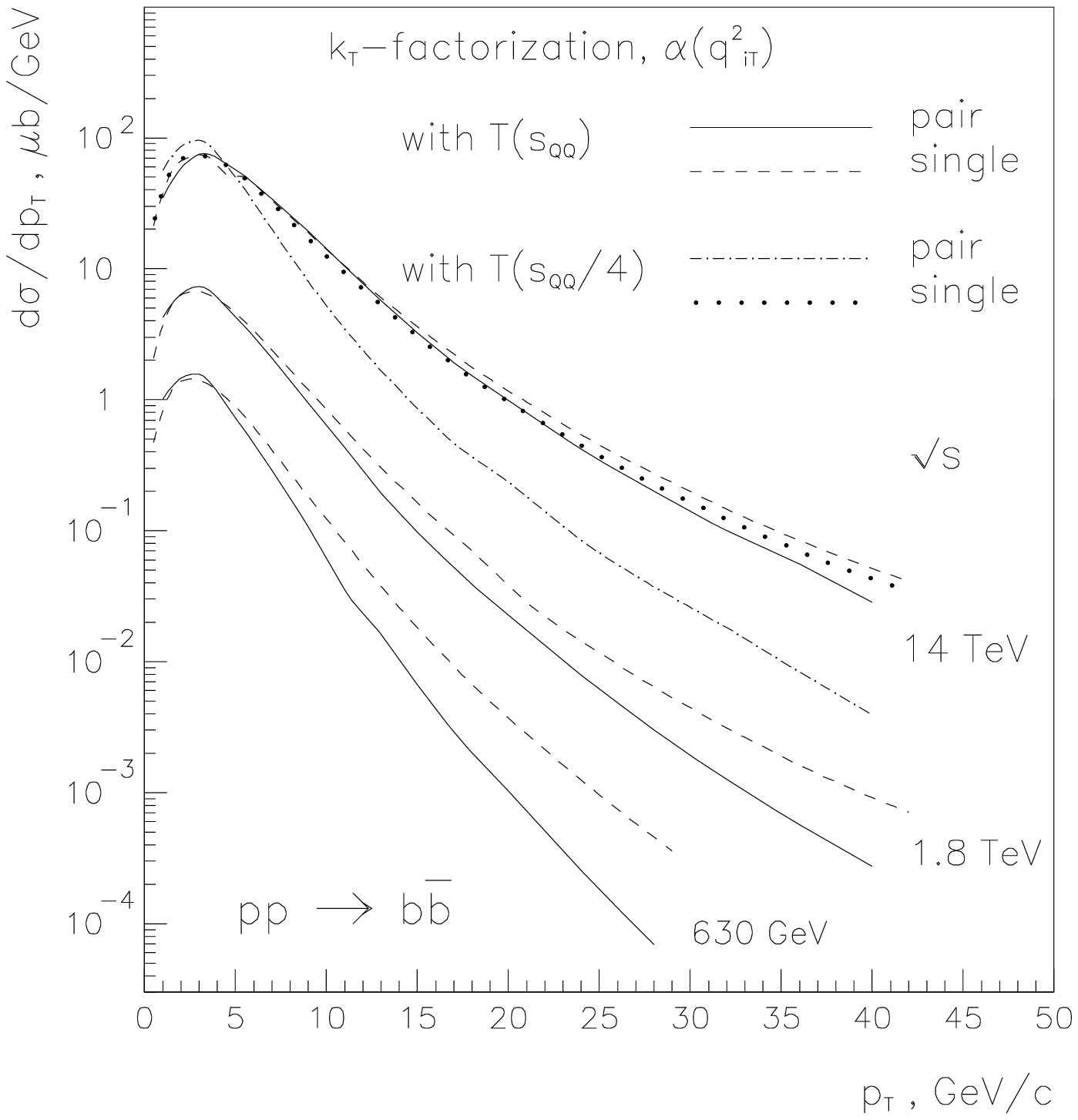,width=0.50\textwidth,clip=}}
Fig. 15. The distributions of the total transverse momentum $p_{pair}$
for $c$-quarks (a) and $b$-quarks (b) produced at different energies
(solid curves), calculated with the unintegrated gluon distribution
$f_g(x,q_T,\mu)$ given by Eq.~(14) and~(17), for $\mu^2$ values in
Eq.~(14) equal to $\hat{s}$. Dashed curves show the one-particle
(single) $p_T$-distributions with the same $\mu^2$ taken from Fig.~13.
Dash-dotted and dotted curves are the same calculations for pair and
single production at $\sqrt{s}$ = 14 TeV with $\mu^2 = \hat{s}/4$.
\end{center}
\end{figure}

At comparatively small energies, $\sqrt{s}$ = 39~GeV and even at
$\sqrt{s}$ =630~GeV the distributions $d \sigma /d p_{pair}$ are
narrower than the one-particle distributions $d \sigma /d p_T$. This
means that the transverse momenta of the produced heavy quarks almost
completely compensate each other. However the situation changes
drastically with increasing of the initial energy. Starting from
comparatively small $p_T$, the difference between the curves decreases
with energy. At $\sqrt{s}$ = 14 TeV the distributions are similar
both in the cases of $c\bar{c}$ and $b\bar{b}$ production. This means
that the production mechanism changes in the discussed energy region.
At $\sqrt{s}$ = 14~TeV the transverse momentum of the produced heavy
quark is balanced more probably by one or several gluons, because the
contribution with large virtuality in the quark propagator is more
suppressed in comparison with the large virtuality in the gluon
propagator.

The discussed behaviour depends on the value of the scale $\mu^2$ in the
$T$-factor, Eq.~(14). The similar calculation at energy $\sqrt{s}$
= 14~TeV with $\mu^2 = \hat{s}/4$ is shown in Fig.~15 for pair and
single production by dash-dotted and dotted curves, respectively.
Here the difference between these two curves is more significant and
becomes larger for lower energies.

The distributions of the produced heavy quark pair as a function of
the rapidity gap $\Delta y = \vert y_Q - y_{\bar{Q}}\vert$ between
quarks are presented in Figs.~16. Difference between the LO
PM and the $k_T$-factorization predictions is not large again except
for the region of very large $\Delta y$.

\begin{figure}
\begin{center}
\centerline{\epsfig{figure=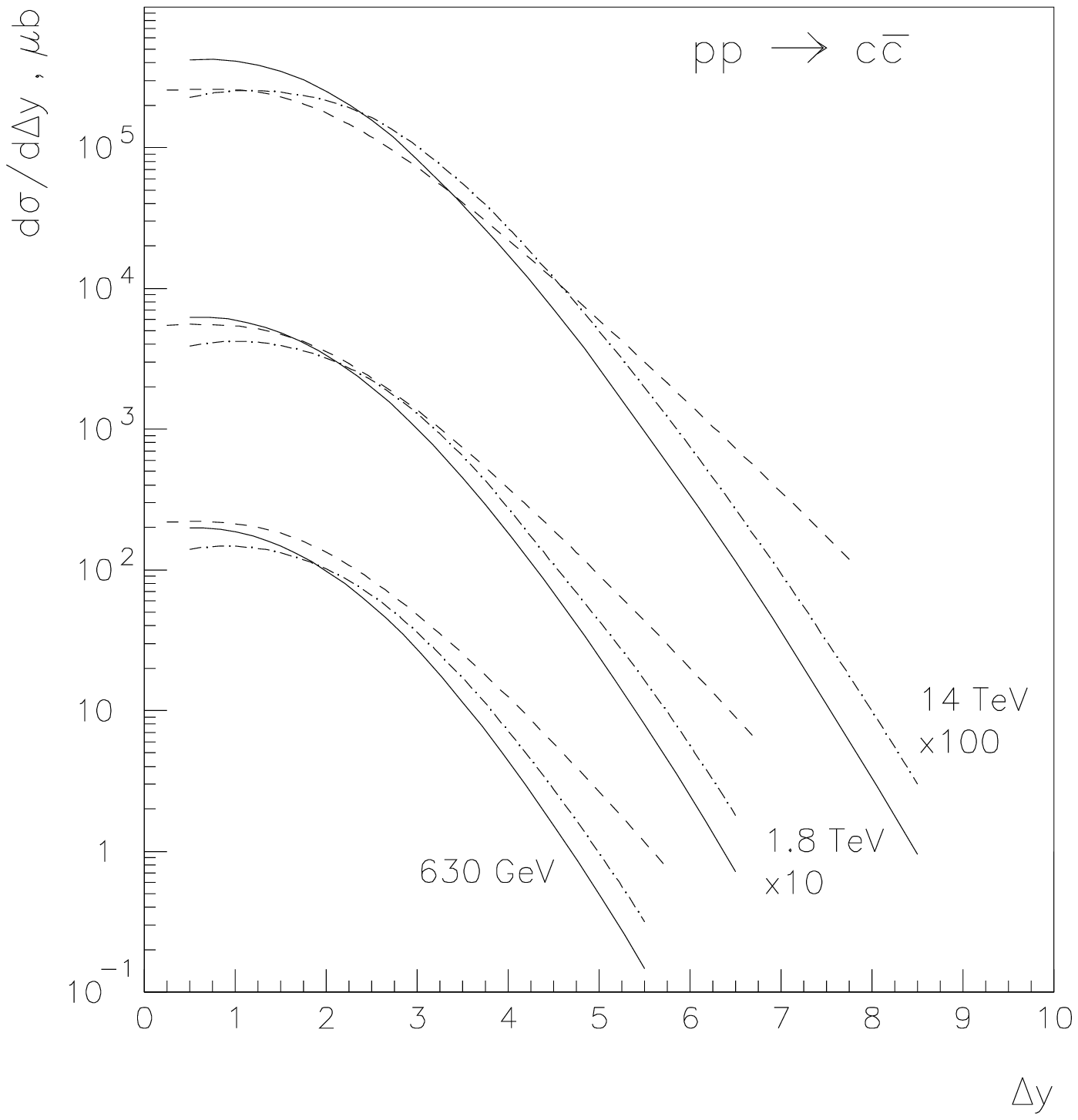,width=0.50\textwidth,clip=}
            \hspace{0.3cm}
            \epsfig{figure=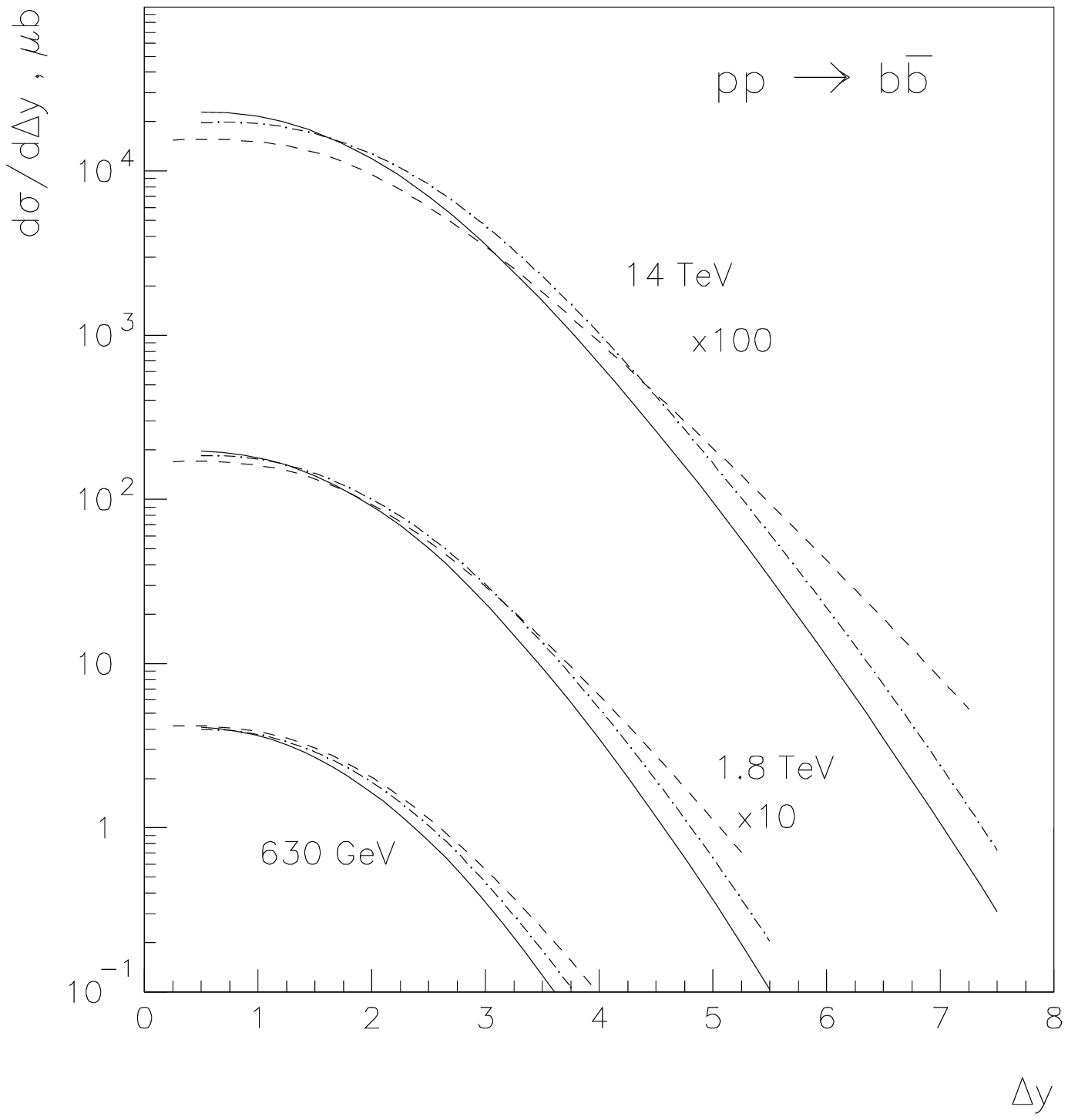,width=0.50\textwidth,clip=}}
Fig. 16. The calculated distributions of the rapidity difference between
two $c$-quarks (a) or $b$-quarks (b) produced at different energies in the
$k_T$ factorisation approach, calculated with unintegrated gluon
distribution $f_g(x,q_T,\mu)$ given by Eq.~(17), for $\mu^2$ values
in Eq.~(14) equal to $\hat{s}$ (solid curves) and $\mu^2 = \hat{s}/4$
(dash-dotted curves). Dashed curves show the LO PM predictioms.
\end{center}
\end{figure}

Another interesting correlation is the distribution of the azimuthal
angle $\phi$. Preliminary results for the azimuthal correlations in
the $k_T$-factorization approach were considered in \cite{SS}. The
main difference in the information coming from $d \sigma /d p_{pair}$
and $d \sigma/d \phi$ distributions is due to the comparatively slow
heavy quark. It gives a negligibly small contribution to the $d
\sigma /d p_{pair}$ in the first case, whereas in the second one each
quark contributes to the distribution $d \sigma/d \phi$ practically
independently of its momentum, so all corrections coming from quark
confinement, hadronization and resonance decay can be important.

As was discussed above, the first contribution in Eq.~(35) is the same
as the conventional LO parton model in which the angle between the
produced heavy quarks is always 180$^o$. However the angle between two
heavy hadrons can be slightly different from this value due to
hadronization processes. To take this into account we assume that in
this first contribution the probability to find a hadron pair with
azimuthal angle $\phi = 180^o -\phi_1$ is determined by the expression
\begin{equation}
w_1(\phi_1)\,=\,\frac{p_h}{\sqrt{p_h^2+p_T^2}} \;,
\end{equation}
where $p_h=0.2$~Gev/c is a transverse momentum in the azimuthal plane
coming from the hadronization process. The other contributions of
Eq.~(35) result in a more or less broad $\phi$ distribution so we
neglect their small modification due to hadronization.

The $k_T$-factorization approach predictions for the azimuthal
correlation of heavy quarks produced in $pp$ collisions are presented
in Fig.~17. One can see that they change drastically when the
initial energy increases from fixed target to the collider region.

\begin{figure}
\begin{center}
\centerline{\epsfig{figure=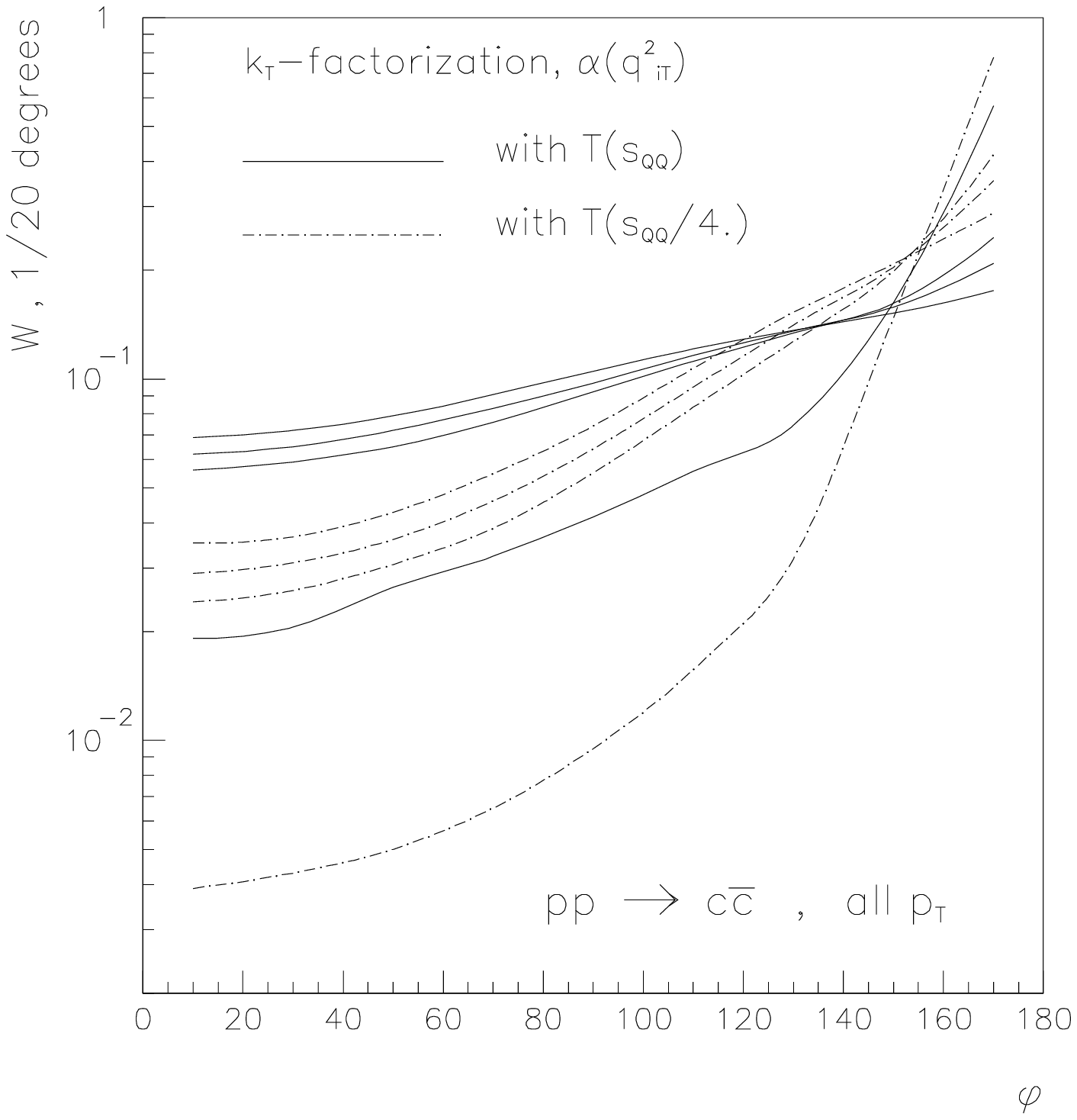,width=0.50\textwidth,clip=}
            \hspace{0.3cm}
            \epsfig{figure=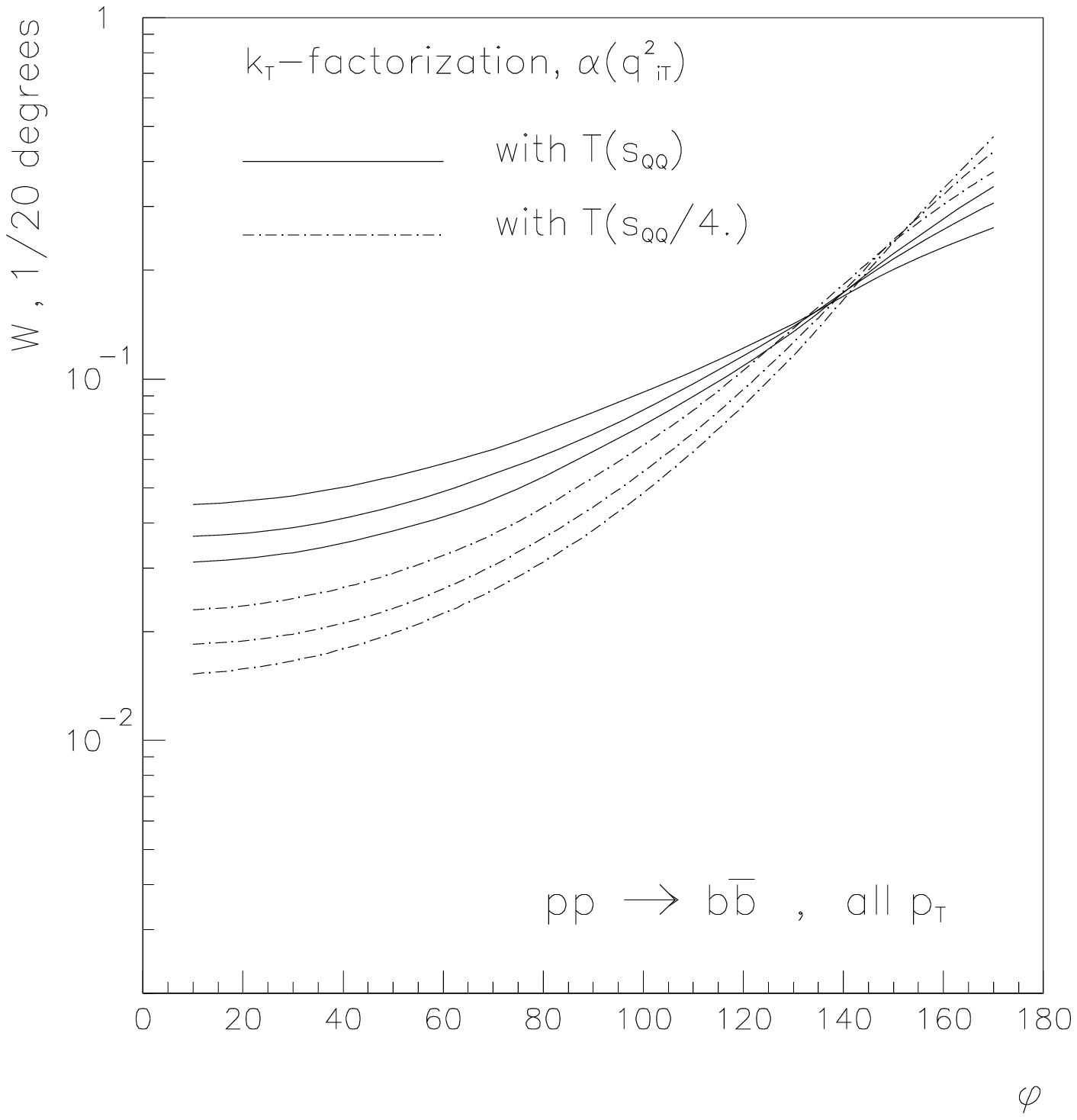,width=0.50\textwidth,clip=}}
Fig. 17. The calculated azimuthal correlations of charm (a) and beauty
(b) pair production for $\mu^2=\hat{s}$ (solid curves) and
$\mu^2=\hat{s}/4$ (dash-dotted curves) at the energies equal to
$\sqrt{s}$ = 14 TeV (upper curves at small $\phi$), 1.8 TeV, 630 GeV
and 39 GeV (the latter one only for charm production).
\end{center}
\end{figure}

There exists a lot of various correlations which have never been
considered both theoretically and experimentally. Let us discuss
several of them. In Fig.~18 we present the distribution on heavy
quark pair transverse momentum in events, when the transverse
momentum of one quark is fixed in the interval $p_{1T}$ = 19-21~
GeV/c. These distributions are very sensitive to the scale value
$\mu^2$ in Eqs.~(14), (17). In the case of $\mu^2 = \hat{s}$ both
the charm and beauty distributions are practically flat for
$p_{pair} < p_{1T}$ (with evident exception for the kinematical
minimum at very small $p_{pair}$) and decrease rather fast for
$p_{pair} > p_{1T}$.  Such a behaviour again shows that the high
transverse momentum of one heavy quark can be compensated by the second
quark (region of comparatively small $p_{pair}$ as well as by the hard
gluon (region of $p_{pair} \sim p_{1T}$). For the smaller
scale $\hat{s}/4$ in Eqs.~(14), (17) the emission of hard gluons is
suppressed and the dashed curves decrease immediately after the
kinematical minima. Thus we can conclude that the measurement of this
distribution allows to find the most reasonable scale value.

\begin{figure}
\begin{center}
\centerline{\epsfig{figure=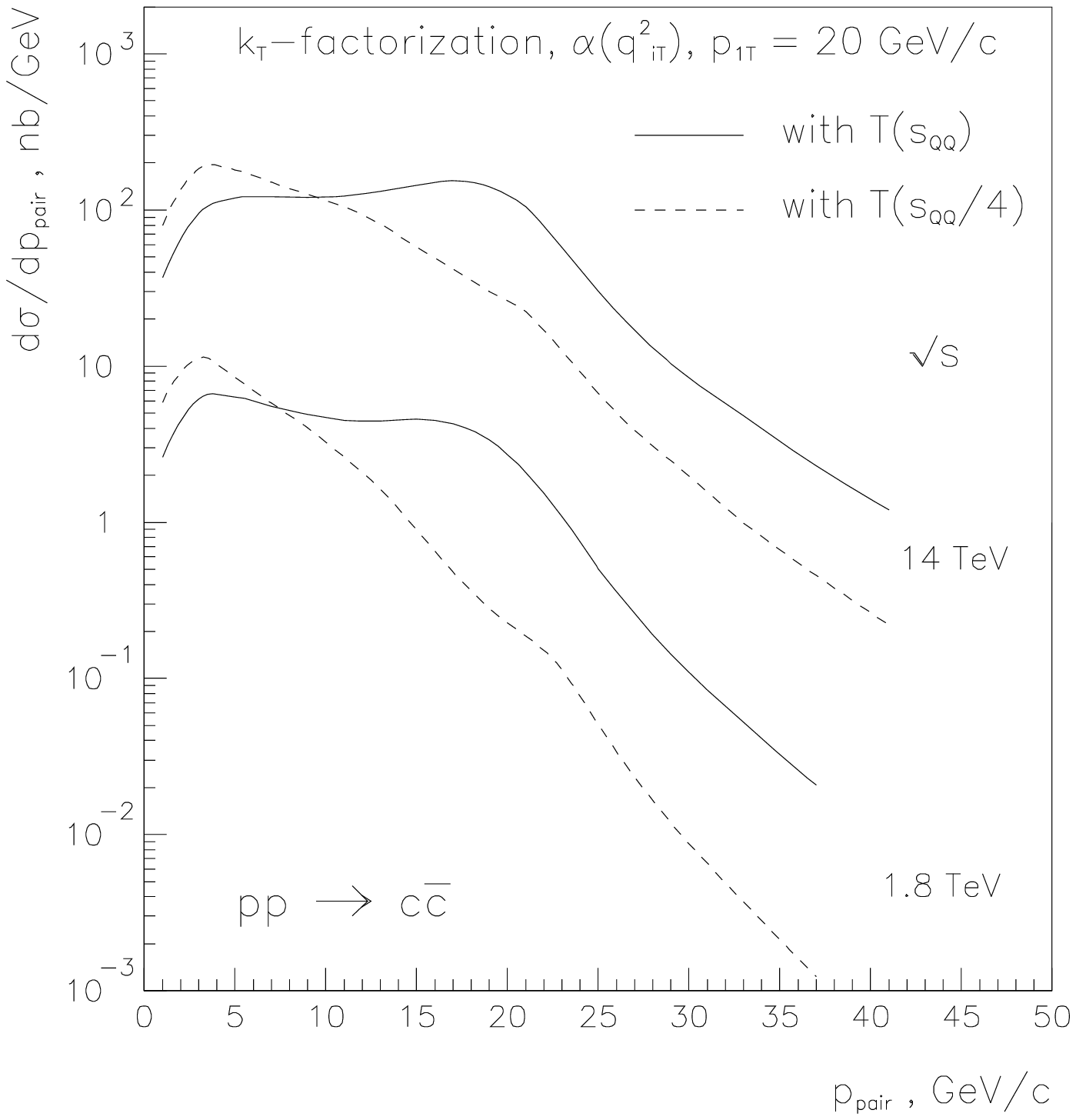,width=0.50\textwidth,clip=}
            \hspace{0.3cm}
            \epsfig{figure=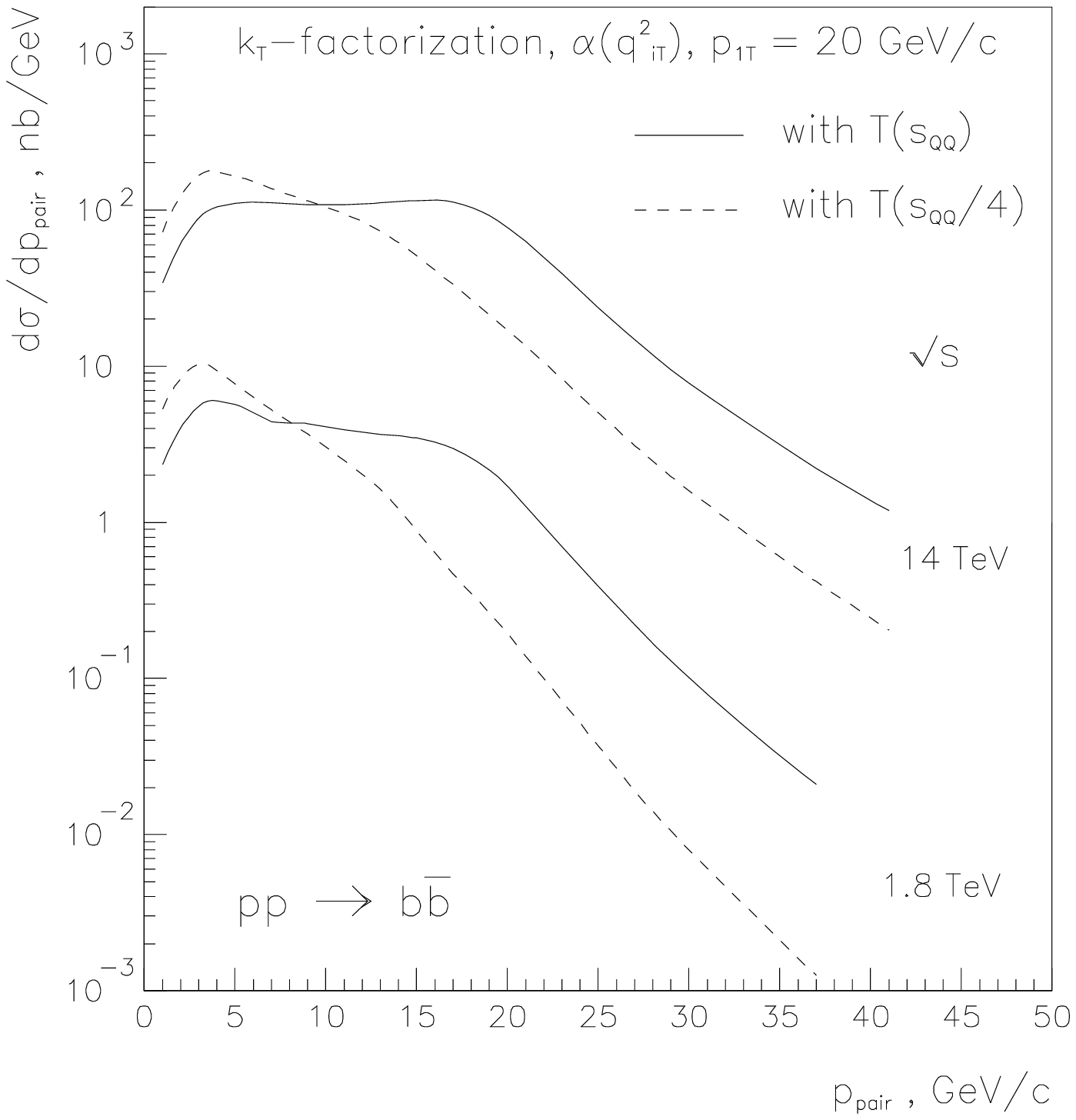,width=0.50\textwidth,clip=}}
Fig. 18. The distributions of the total transverse momentum $p_{pair}$
for $c$-quarks (a) and $b$-quarks (b) produced at different energies
(solid curves) and with the transverse momentum of a one heavy quark
restricted to the interval 19-21 GeV/c, calculated with the
unintegrated gluon distribution $f_g(x,q_T,\mu)$ given by Eq.~(14)
and~(17), for $\mu^2$ values in Eq.~(14) equal to $\hat{s}$.  Dashed
curves show the same distributions with scale $\mu^2 = \hat{s}/4$.
\end{center}
\end{figure}

The scale problem can be solved if we will consider some
distributions in different azimuthal angle regions. For example, the
one-dimentional transverse momentum distribution is insensitive to
the scale for back-to-back production ($\Delta \phi = 120^o - 180^o$)
and rather sensitive for the smaller azimuthal angles, see Figs.~19 and
20.

\begin{figure}
\begin{center}
\centerline{\epsfig{figure=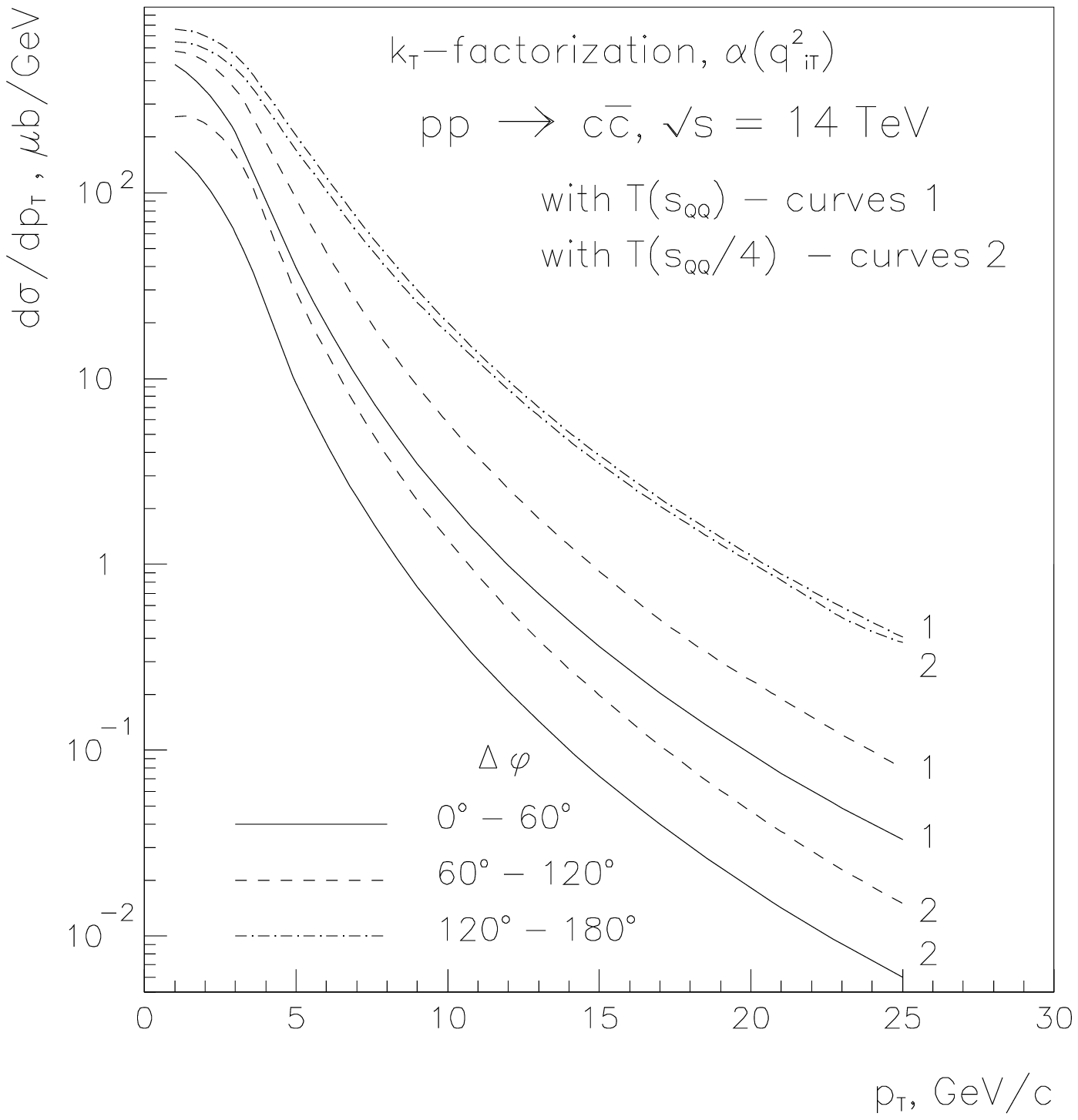,width=0.50\textwidth,clip=}
            \hspace{0.3cm}
            \epsfig{figure=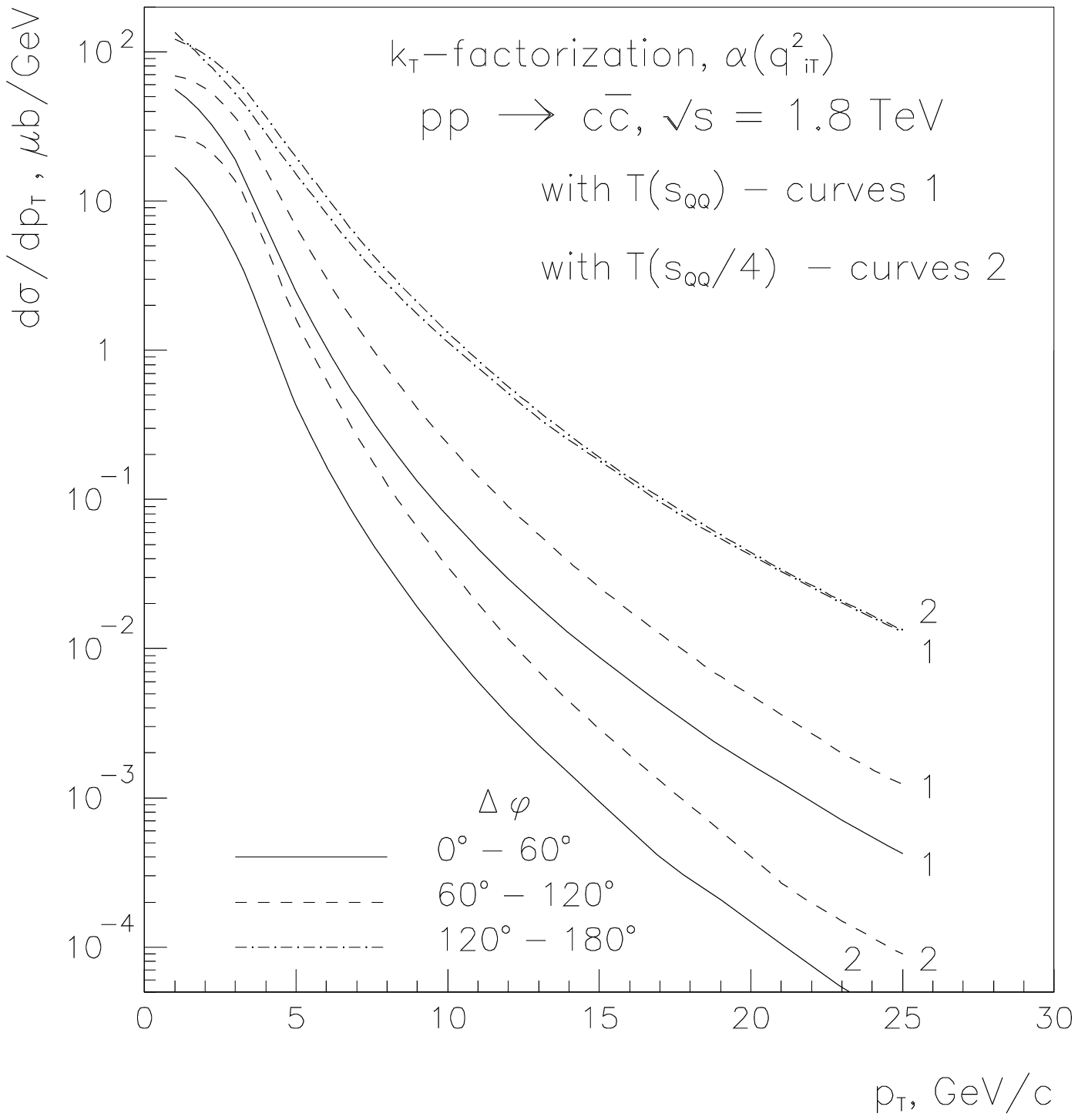,width=0.50\textwidth,clip=}}
Fig. 19. $p_T$-distributions of $c$-quarks in different azimuthal
angles produced at different energies.
\end{center}
\end{figure}

\begin{figure}
\begin{center}
\centerline{\epsfig{figure=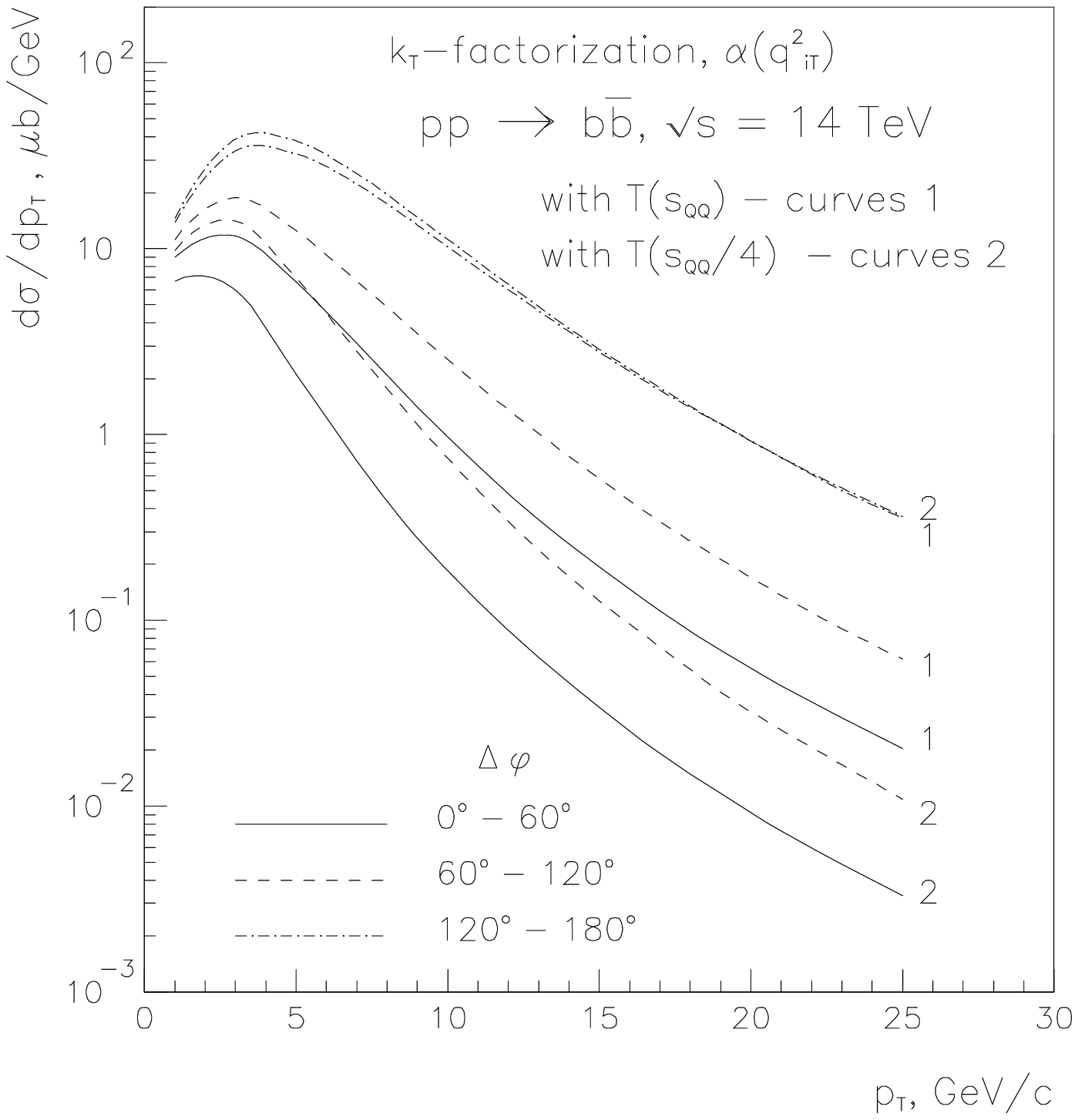,width=0.50\textwidth,clip=}
            \hspace{0.3cm}
            \epsfig{figure=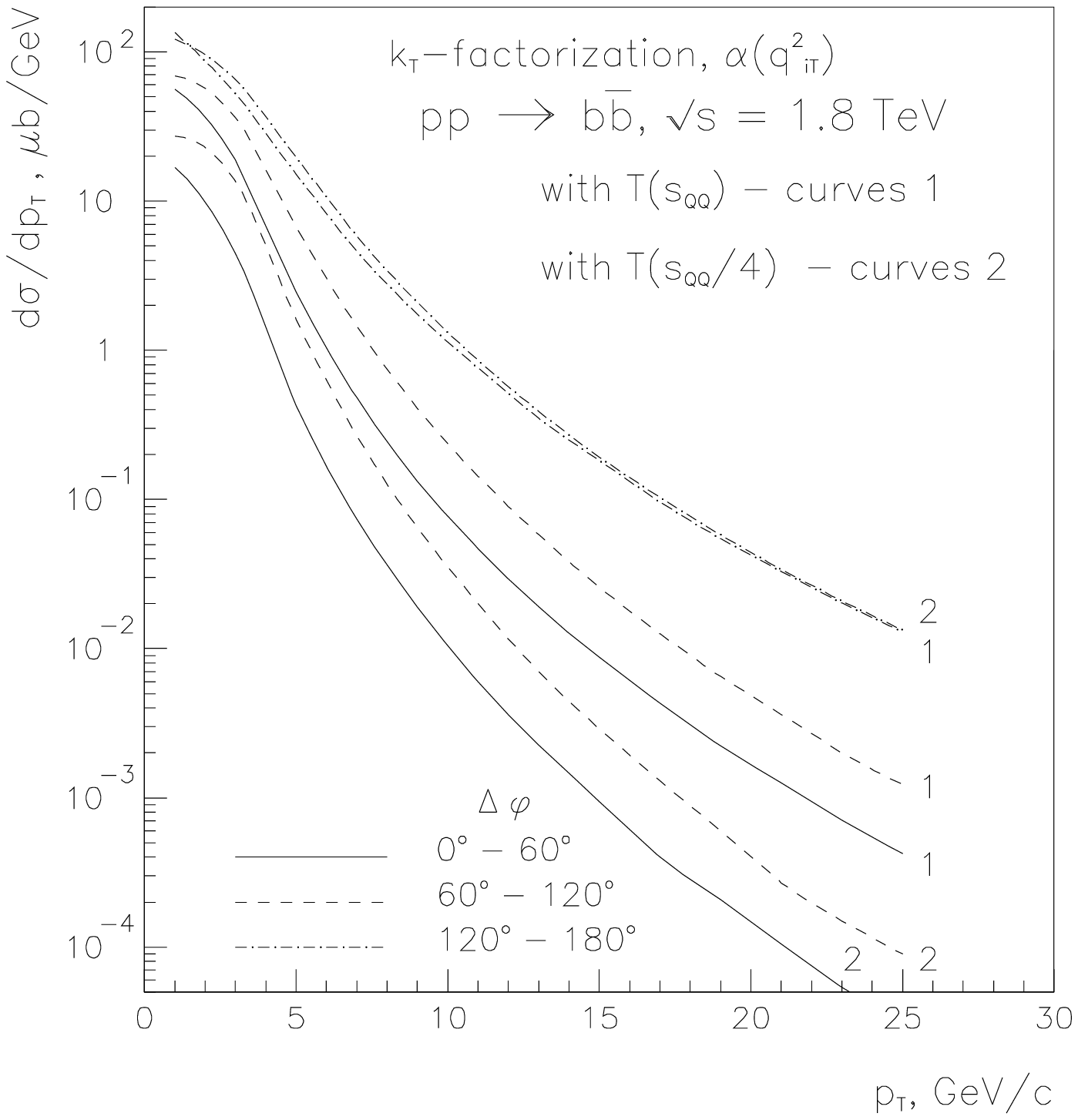,width=0.50\textwidth,clip=}}
Fig. 20. $p_T$-distributions of $b$-quarks in different azimuthal
angles produced at different energies.
\end{center}
\end{figure}

The distribution over $\Delta y$ at LHC energy is predicted to have
minimum at the small azimuthal angles and $\Delta y$ values in the case
of comparatively small scale $\mu^2 = \hat{s}/4$ and to be more flat
for the larger scale, see Figs.~21, 22.

\begin{figure}
\begin{center}
\centerline{\epsfig{figure=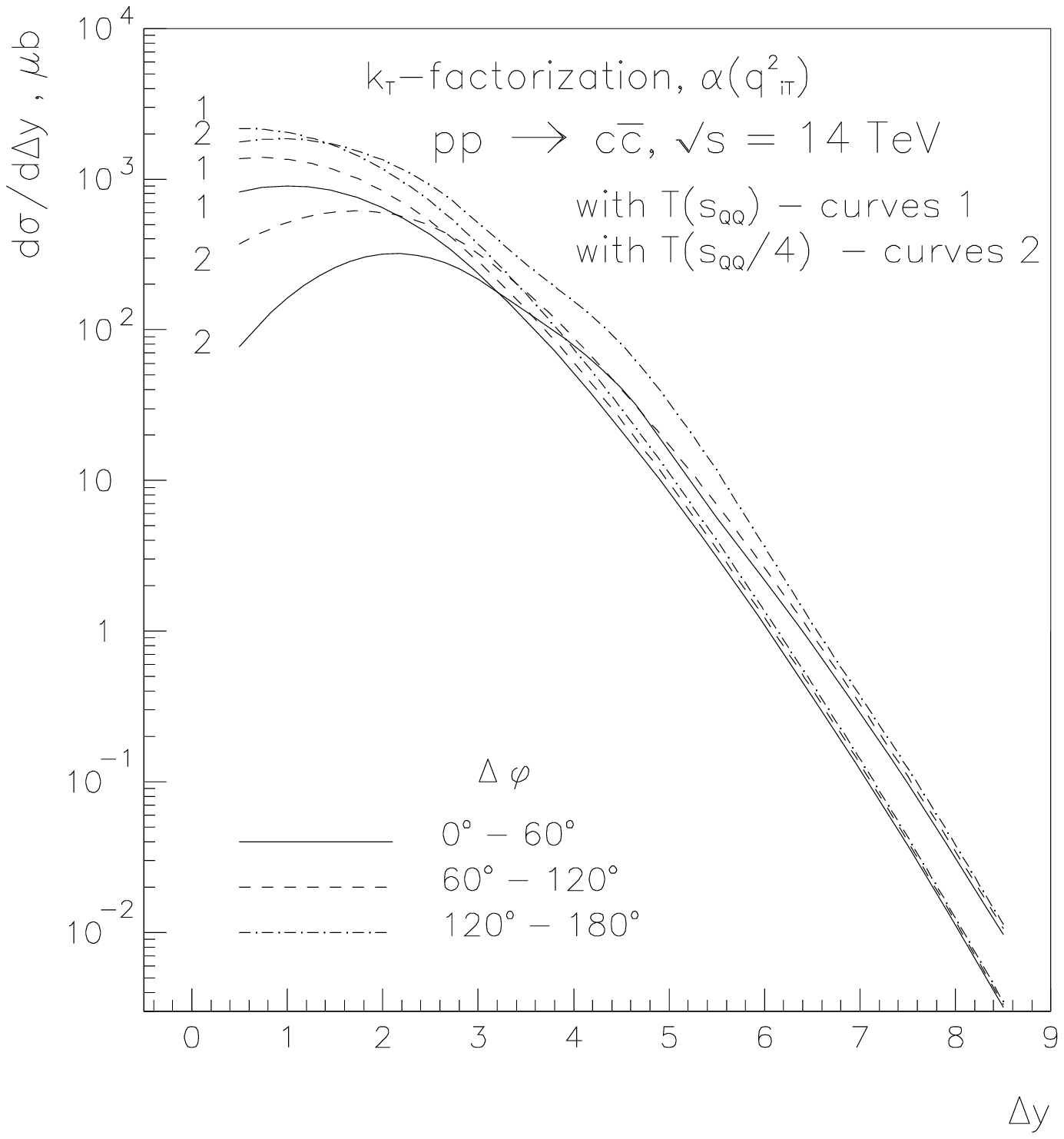,width=0.50\textwidth,clip=}
            \hspace{0.3cm}
            \epsfig{figure=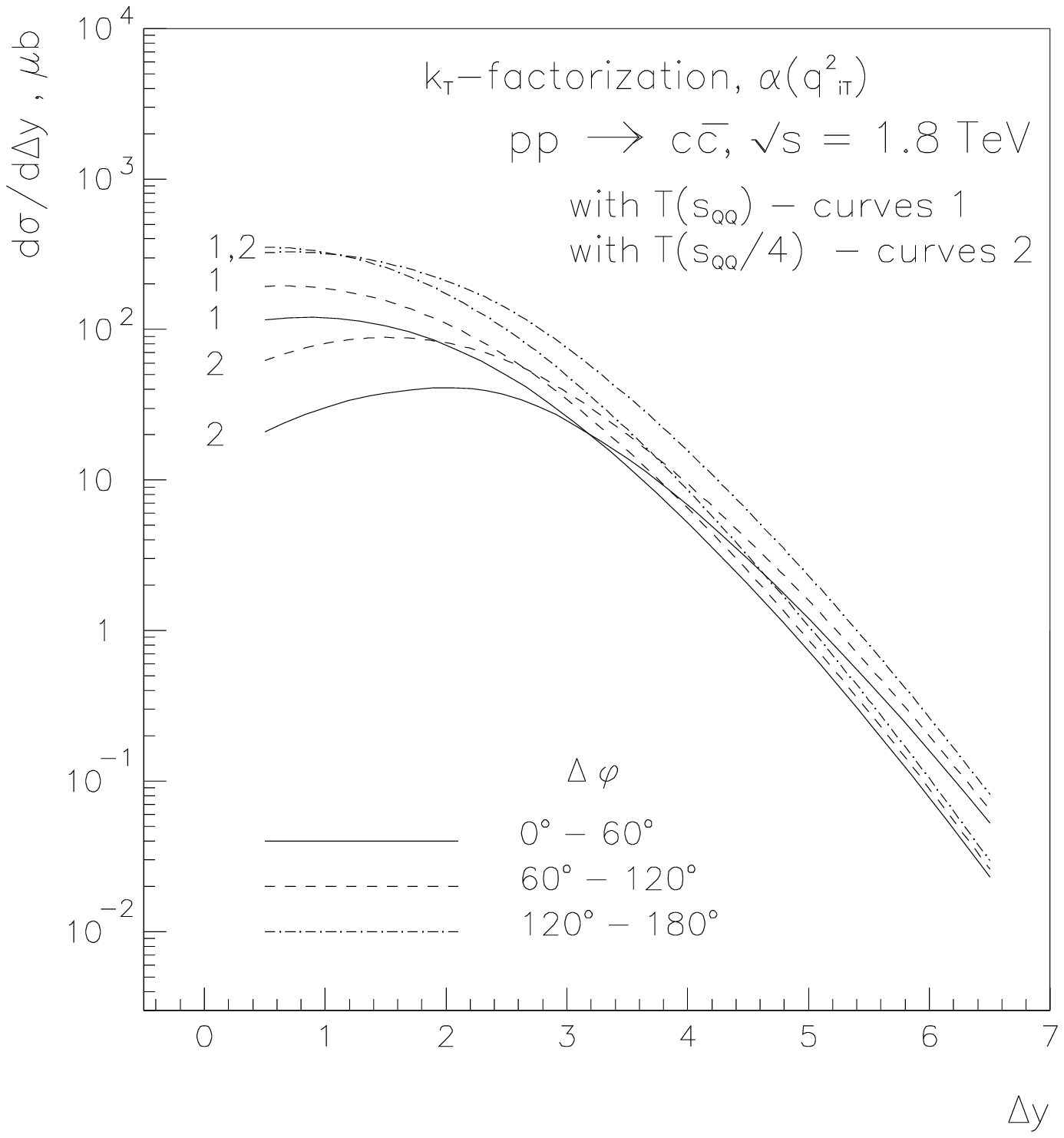,width=0.50\textwidth,clip=}}
Fig. 21. The calculated distributions of the rapidity difference
between two $c$-quarks produced at different azimuthal angles in the
$k_T$ factorisation approach.
\end{center}
\end{figure}

\begin{figure}
\begin{center}
\centerline{\epsfig{figure=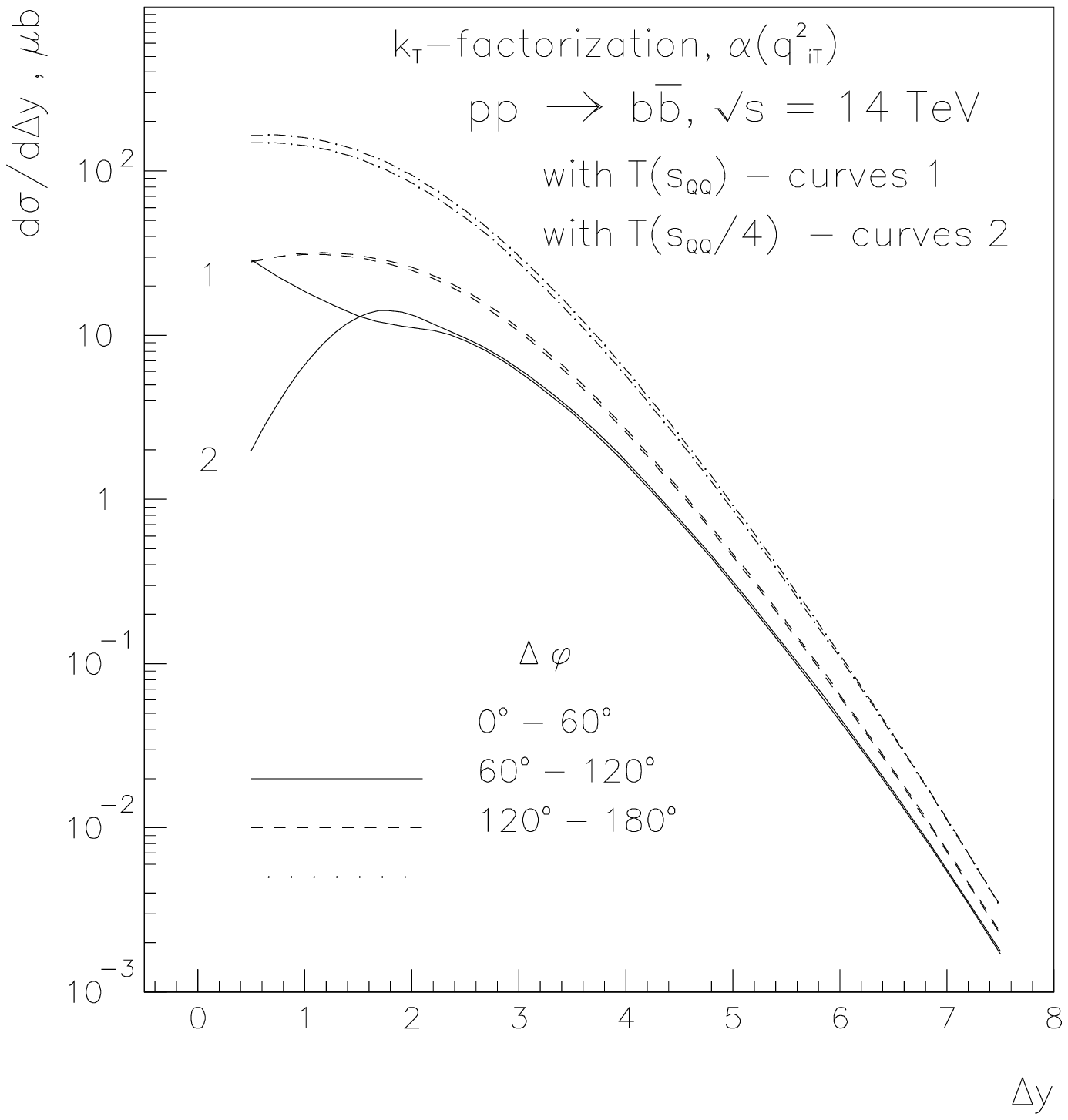,width=0.50\textwidth,clip=}
            \hspace{0.3cm}
            \epsfig{figure=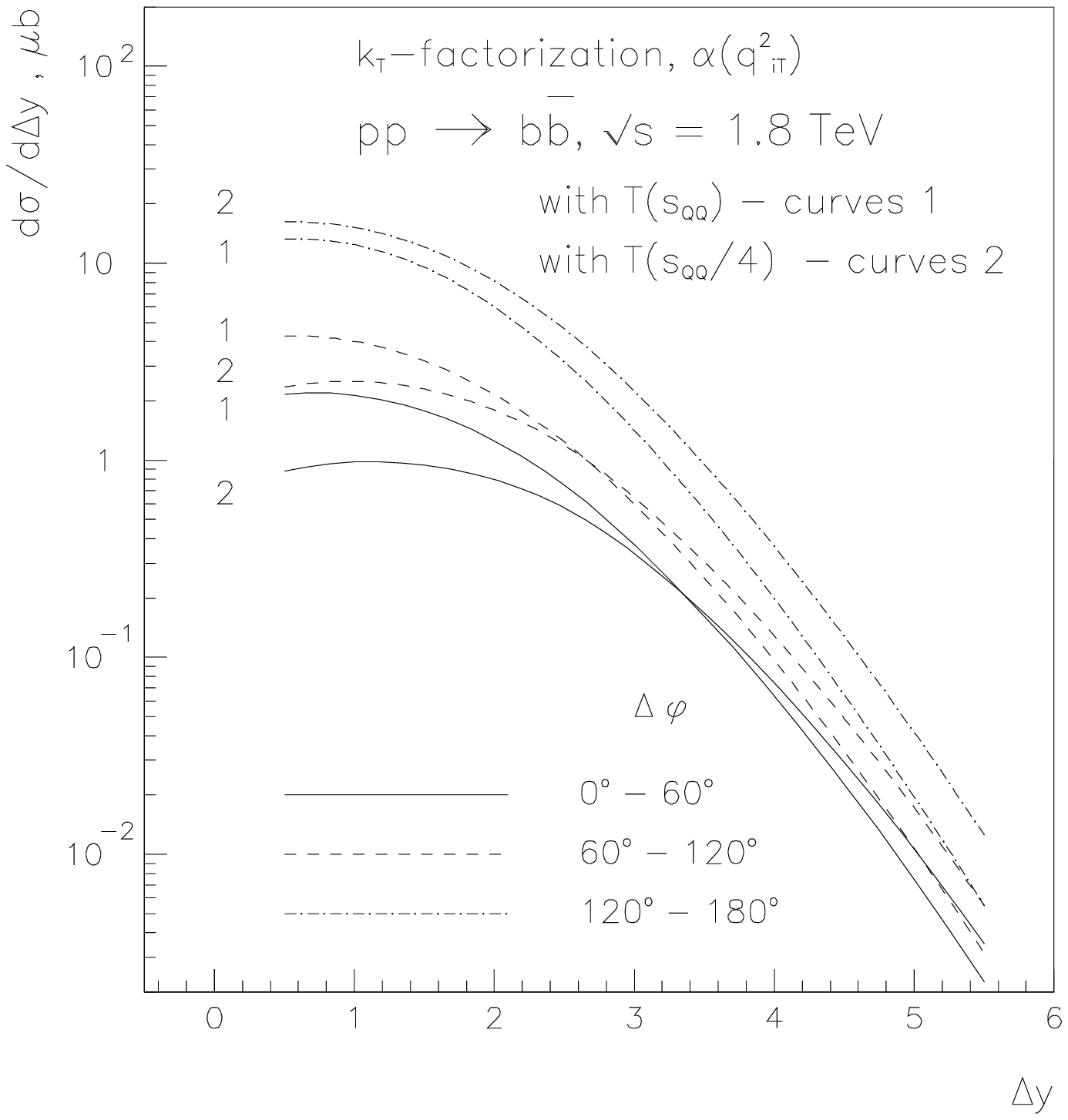,width=0.50\textwidth,clip=}}
Fig. 22. The calculated distributions of the rapidity difference
between two $b$-quarks produced at different azimuthal angles in the
$k_T$ factorisation approach.
\end{center}
\end{figure}

\section{Conclusion}

We have compared the conventional LO Parton Model (PM) and the
$k_T$-factorization approach for heavy quark hadroproduction at fixed
target and collider energies using both "toy" and realistic gluon
distributions. Both the transverse momenta and rapidity
distributions have been considered, as well as several types of
two-particle correlations, such as the distribution of the rapidity
gap between two heavy quarks, their azimuthal correlations and
distributions of the total transverse momentum of the produced heavy
quark pair, $p_{pair}$.

It has been shown in Sect.~4 \cite{our} that the contribution of the
domain with strong $q_T$ ordering
($q_{1T}, q_{2T} \ll m_T=\sqrt{m_Q^2+p_T^2}$) coincides in the
$k_T$-factorization approach with the LO PM prediction. Besides this
a numerically large contribution appears at high energies in
$k_T$-factorization approach in the region $q_{1,2T}\ge m_T$. This
kinematically relates to the events where the transverse momentum of
heavy quark $Q$ is balanced not by the momentum of antiquark
$\overline Q$ but by the momentum of the nearest gluon.

This configuration is associated with the NLO (or even NNLO, if both
$q_{1T}, q_{2T} \ge m_T$) corrections in terms of the PM with fixed
number of flavours, i.e. without the heavy quarks in the evolution.
Indeed, as was mentioned in \cite{1}, up to 80\% of the whole NLO
cross section originates from the events where the heavy quark
transverse momentum is balanced by the nearest gluon jet. Thus the
large "NLO" contribution, especially at large $p_T$, is explained by
the fact that the virtuality of the $t$-channel (or $u$-channel)
quark becomes small in the region $q_T \simeq p_T$, and the
singularity of the quark propagator $1/(\hat{p} - \hat{q} - m_Q)$ in
the "hard" QCD matrix element, $M(q_{1T},q_{2T},p_{1T},p_{2T})$,
reveals itself.

The double logarithmic Sudakov-type form factor $T$ in the definition
of the unintegrated parton density Eq~(17) comprises an important part of
the virtual loop NLO (with respect to the PM) corrections. Thus we
demonstrate that the $k_T$-factorization approach collects already at the
LO the major part of the contributions which play the role of the NLO
(and even NNLO) corrections to the conventional PM.  Therefore we
hope that a higher order (in $\alpha_S$) correction to the
$k_T$-factorization could be rather small.

Another advantage of this approach is that a non-zero transverse
momentum of $Q \overline Q$-system
($p_{T pair}=p_{1T}+p_{2T} = q_{1T}+q_{2T}$) is naturally achieved in the
$k_T$-factorization. We have calculated the $p_{T pair}$ distribution
and compared it with the single quark $p_T$ spectrum. At the
low energies the typical values of $p_{T pair}$ are much lesser than the heavy
quark $p_T$ in accordance with collinear approximation. However
for LHC energy both spectra become close to each other indicating that
the transverse momentum of second heavy quark is relatively small. The
typical value of this momentum ($p_{T pair}=k_T$-kick) depends on the
parton structure functions/densities. It increases with the initial
energy ($k_T$-kick increases with the decreasing of the momentum
fractions $x,y$ carried by the incoming partons) and with the
transverse momenta of the heavy quarks, $p_T$. Thus one gets a possibility
to describe a non-trivial azimuthal correlation without introducing a
large "phenomenological" intrinsic transverse momentum of the partons.

It is necessary to note that the typical parton
transverse momenta $q_{1T}$ and $q_{2T}$ increase in our calculations
with the growth of the detected $b$-quark momentum $p_T^{min}$. In
the language of $k_T$-kick it means that the values of
$\langle k_T^2\rangle$ also increase.

At the time being we would not like to compare our
calculations with experimental data\footnote{An example of very
successful comparison in the $k_T$-factorization approach can be found in
\cite{HKSST} where some different assumptions were used.}. It is
shown that in all cases when a realistic gluon
distribution is used the difference between our approach and the conventional
parton model is not large, so our results should be in agreement with the
data up to, say, a factor 2 or 3. At the same time
the disagreement of such an order seems to be senseless to discuss,
because the unintegrated gluon distributions are not known with the needed
accuracy (from the evolution equation and the experimental data).
Instead of them we have used more or less reasonable approximation Eq.~(17).
Although it is qualitatively good it can lead to a numerical
disagreement. In particular, the angular ordering \cite{MCi,CFM,Mar}
implies that the cut-off $\Delta = q_T/\mu$ in Eqs.~(14), (17) should
be replaced by $\Delta = q_T/(\mu +q_T)$.
From the formal point of view the difference is beyond the DGLAP LO
accuracy. However with the new $\Delta = q_T/(\mu +q_T)$ one gets the
non-zero values of $f_a(x, q_T, \mu)$ even at the large $q_T > \mu$. This
is especially important at the high energies where the essential
values of $x$ and $z$ (in Eq.~(17)) are very small.
The contribution coming from $q_T > \mu$ region could enhance the
flux of colliding gluons up to a few times, and by this way explain
the FNAL-Tevatron puzzle. These new data on the cross section of
$b\bar{b}$ (or high-$p_T$ prompt photon) production are 2-3 times
larger than the conventional NLO PM QCD predictions \cite{Abb,Apan,KMR}.
At the moment we have no realistic unintegrated parton distributions
which fit the data with accounting for the contribution from
$q_T > \mu$.

That is why the presented here results are treated mainly as the
qualitatively ones, having the numerical accuracy on the level of
factor 2-3.

Another important point is that almost
all presented results concern the heavy quarks rather than the hadron
production.
Of course, the hadronization leads to several important
effects, such as different yields (asymmetry) of $D$- and
$\bar{D}$-mesons production in $pp$ collisions, see qualitative
discussion in \cite{YMSh}, however their quantitative description
needs additional phenomenological assumptions, see e.g.
\cite{Likh,TNKN,Likh1,jdd} or description of PYTHIA Monte Carlo code.

\subsection*{Acknowledgements}

We are grateful to E.M.Levin who participated at the early stage of
this activity and to many our colleagues for useful discussions.
The presented calculations were carried out in part in ICTP.
One of us (Y.M.S) is grateful to Prof. S.Randjbar-Daemi for providing
this possibility and to the staff for creating good working
conditions. We are grateful to Yu.L.Dokshitzer, G.P.Korchemsky and
M.N.Mangano for discussions.\\
This work was supported by grants NATO OUTR.LG 971390 and RFBR
98-02-17629.

%\newpage

\end{document}